\documentclass{article}

\usepackage{microtype}
\usepackage{graphicx}
\usepackage{subcaption}
\usepackage{booktabs} 

\usepackage{hyperref}

\usepackage{tcolorbox}
\tcbuselibrary{breakable, skins}

\newtcolorbox{promptbox}[2][]{%
    colback=gray!5!white,      %
    colframe=gray!50!black,    %
    colbacktitle=gray!70!black,%
    coltitle=white,            %
    fonttitle=\bfseries,       %
    title={#2},                %
    arc=1mm,
    boxrule=0.5mm,
    left=2mm,
    right=2mm,
    top=2mm,
    bottom=2mm,
    breakable,                 %
    enhanced,
    fontupper=\small,
    #1                         %
}                              %

\newcommand{\placeholder}[1]{\texttt{\{#1\}}}

\usepackage[preprint]{icml2026}

\usepackage{amsmath}
\usepackage{amssymb}
\usepackage{mathtools}
\usepackage{amsthm}
\usepackage{booktabs}
\usepackage{multirow}
\usepackage{amssymb}
\usepackage{graphicx}

\newcommand{\newcheck}{\checkmark}
\newcommand{\newcross}{\ensuremath{\times}}

\newcommand{\y}{\newcheck}
\newcommand{\n}{\newcross}
\usepackage[capitalize,noabbrev]{cleveref}

\newcommand{\model}{\textsc{VenusRAR}}

\newcommand{\ma}{\textit{Mutant A}}
\newcommand{\mb}{\textit{Mutant B}}

\newcommand{\code}{\url{https://github.com/ai4protein/VenusRAR/}}

\theoremstyle{plain}

\theoremstyle{definition}

\theoremstyle{remark}

\usepackage[textsize=tiny]{todonotes}

\icmltitlerunning{Submission and Formatting Instructions for ICML 2026}

\begin{document}

\twocolumn[
  \icmltitle{Rank-and-Reason: Multi-Agent Collaboration Accelerates Zero-Shot\\ Protein Mutation Prediction}
  \icmlsetsymbol{equal}{*}

  \begin{icmlauthorlist}
    \icmlauthor{Yang Tan}{sjtu,sii}
    \icmlauthor{Yuanxi Yu}{sjtu,equal}
    \icmlauthor{Can Wu}{ecust}
    \icmlauthor{Bozitao Zhong}{sjtu}
    \icmlauthor{Mingchen Li}{sjtu}
    \icmlauthor{Guisheng Fan}{ecust}
    \icmlauthor{Jiankang Zhu}{sustech}
    \icmlauthor{Yafeng Liang}{sustech}
    \icmlauthor{Nanqing Dong}{sii}
    \icmlauthor{Liang Hong}{sjtu}
  \end{icmlauthorlist}

  \icmlaffiliation{sjtu}{Shanghai Jiao Tong University}
  \icmlaffiliation{sii}{Shanghai Innovation Institute}
  \icmlaffiliation{sustech}{Southern University of Science and Technology}
  \icmlaffiliation{ecust}{East China University of Science and Technology}

  \icmlcorrespondingauthor{Yafeng Liang}{liangyf@mail.sustech.edu.cn}
  \icmlcorrespondingauthor{Nanqing Dong}{nanqing.dong@sii.edu.cn}
  \icmlcorrespondingauthor{Liang Hong}{hongl3liang@sjtu.edu.cn}

  \icmlkeywords{Protein Engineering, Multi-Agent, Mutation Prediction}

  \vskip 0.3in
]



\printAffiliationsAndNotice{\icmlEqualContribution}

\begin{abstract}
  Zero-shot mutation prediction is vital for low-resource protein engineering, yet existing protein language models (PLMs) often yield statistically confident results that ignore fundamental biophysical constraints. Currently, selecting candidates for wet-lab validation relies on manual expert auditing of PLM outputs, a process that is inefficient, subjective, and highly dependent on domain expertise. To address this, we propose Rank-and-Reason (\model), a two-stage agentic framework to automate this workflow and maximize expected wet-lab fitness. In the Rank-Stage, a \textit{Computational Expert} and \textit{Virtual Biologist} aggregate a context-aware multi-modal ensemble, establishing a new Spearman correlation record of $0.551$ (vs. $0.518$) on \textsc{ProteinGym}. In the Reason-Stage, an agentic \textit{Expert Panel} employs chain-of-thought reasoning to audit candidates against geometric and structural constraints, improving the Top-5 Hit Rate by up to $367\%$ on \textsc{ProteinGym-DMS99}. The wet-lab validation on \texttt{Cas12i3} nuclease further confirms the framework's efficacy, achieving a $46.7\%$ positive rate and identifying two novel mutants with $4.23$-fold and $5.05$-fold activity improvements. Code and datasets are released on GitHub\footnote{\code}.
\end{abstract}

\section{Introduction}

 Protein engineering presents a grand challenge for science and industry. From designing enzymes for catalysis \cite{wu2025pet} to engineering therapeutic antibodies \cite{chai2025chai2}, optimizing protein function is a critical objective. Most natural proteins are not optimized for industrial applications, necessitating directed engineering to enhance specific properties. While the theoretical sequence space is combinatorial ($20^L$) and computationally intractable to traverse exhaustively, the subspace of single-point mutations remains a tractable and high-value target for exploration \cite{yang2025alde}. A single critical mutation can dramatically enhance the function of a given protein (e.g., \textit{T203Y} mutation that derived EGFP from GFP \cite{heim1996egfp}). Furthermore, identifying robust positive single mutants provides the essential building blocks for subsequent multi-point recombination strategies \cite{zhou2024mlife}. Unlike data labeling in machine learning \cite{snow2008nlp_label_cheap,russakovsky2015imagenet}, validating a single protein mutant requires resource-intensive wet-lab procedures including gene synthesis, expression, and purification \cite{rosano2014weblab_expensive,li2025VenusVaccine}. 
 
 Such reality gives rise to the \textbf{Low-$N$} protein engineering problem: selecting a small number of candidates from the vast mutational landscape for experimental validation. In such budget-constrained scenarios, the success of an engineering is critically associated with the quality of initial candidate selection. Random screening typically yields positive rates below 1\% \cite{romero2009exploring}, making intelligent zero-shot prediction—ranking candidates before any experimental feedback—the most critical bottleneck in the engineering lifecycle \cite{jiang2024prime}.

To address this challenge, the field has shifted from physics-based methods~\cite{schymkowitz2005foldx,das2008rosetta} to sequence, structure, and hybrid data-driven PLMs \cite{notin2022tranception,zhang2024s3f,tan2025venusrem} pre-trained on vast protein datasets, which now dominate benchmarks such as \textsc{ProteinGym}~\cite{notin2024proteingym} and \textsc{VenusMutHub} \cite{zhang2025venusmuthub}. Modern PLMs integrate sequence, structure, and evolutionary information~\cite{su2023saprot,tan2025protssn,tan2025venusrem}, achieving strong correlations with experimental fitness, and their reliability is increasingly validated by wet-lab experiments \cite{jiang2024prime,jiang2024evolvepro,tan2025venusrem,Zheng2026protssn_wet_lab}.

Although PLMs demonstrate substantial predictive capabilities, they frequently fail to incorporate essential biophysical constraints and lack the explicit interpretability required for scientific applications \cite{gujral2025blakcbox,gelman2025biophysics}. The zero-shot protein engineering domain has long been confined to a tool-centric phase \cite{tan2025venusfactory}, relying on human experts to manually evaluate model outputs to identify viable candidates, lacking the necessary progression toward autonomous scientific discovery. Large Language Models (LLMs) such as \textsc{GPT-4} \cite{achiam2023gpt4} and \textsc{DeepSeek} \cite{guo2025deepseek} can improve the progression by providing a vast scientific knowledge base and an explicit reasoning process. While LLM-based autonomous agents were initially designed for specific tasks within the artificial intelligence domain, such as automated research and manuscript generation \cite{lu2024ai-scientist}, studies have shown that multi-agent collaboration can significantly improve the overall efficiency~\cite{li2024more,su-etal-2025-many}. So far, multi-agent systems have been developed in various scientific fields to support \textit{de novo} antibody design \cite{swanson2025virtuallab,zhou2025prime}, life science discovery \cite{gao2024bio_discovery_agent,zhao2025rare_disease_agent}, and scientific tool integration \cite{m2024chemcrow,ding2025scitoolagent}. Despite these developments, an effective multi-agent framework specifically designed for the constraints of directed evolution is still absent.

\begin{figure*}[ht]
  \begin{center}
    \centerline{\includegraphics[width=\textwidth]{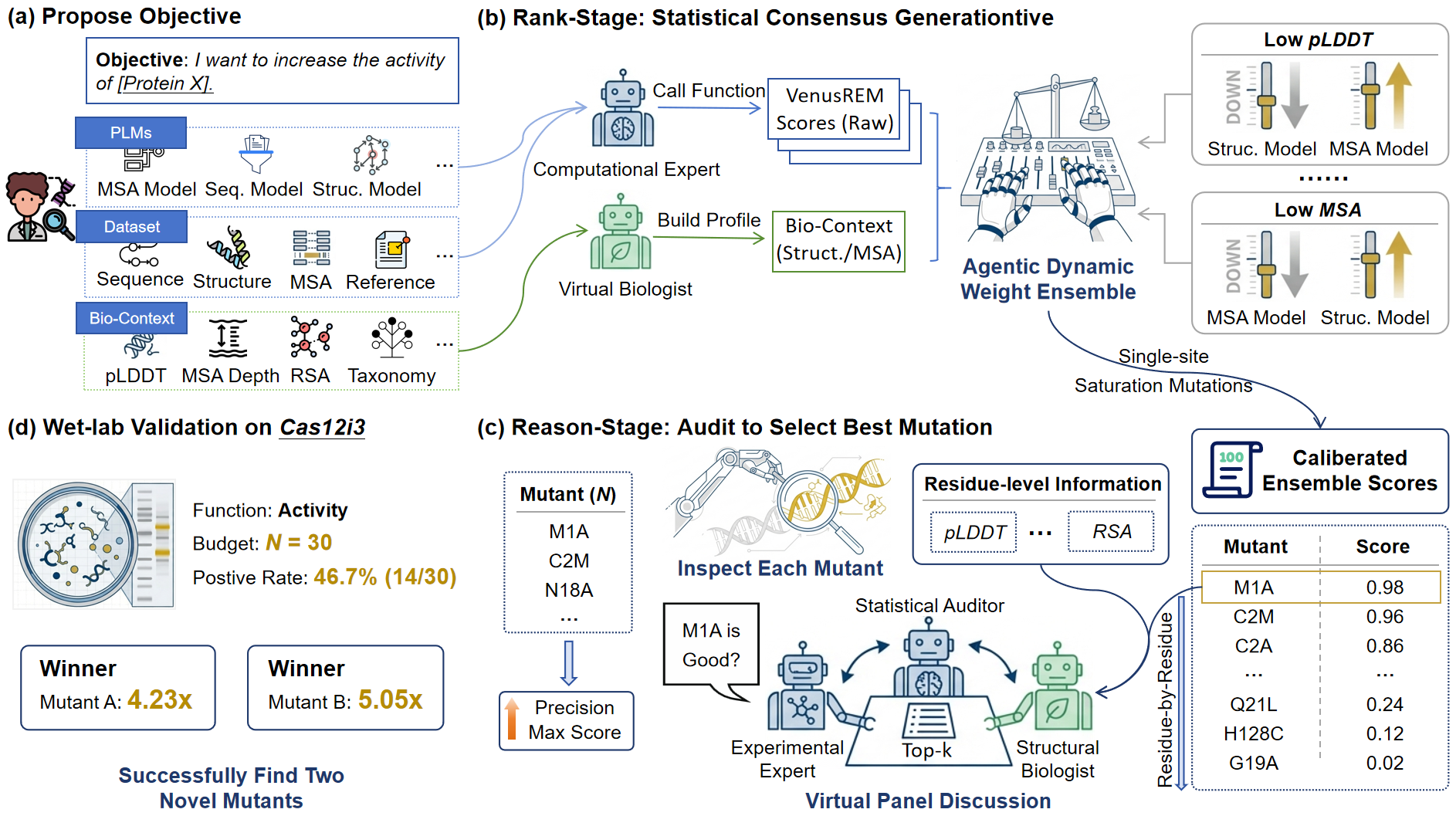}}
    \caption{
      The~\model~multi-agent framework. \textbf{(a)} Users define engineering objectives and multi-modal contexts. \textbf{(b)} \textbf{Rank-Stage}: A \textit{Computational Expert} aggregates raw PLM scores, while a \textit{Virtual Biologist} dynamically calibrates ensemble weights based on structural and evolutionary data quality. \textbf{(c)} \textbf{Reason-Stage}: A \textit{Virtual Expert Panel} conducts CoT-based auditing to filter biophysically invalid candidates. \textbf{(d)} Wet-lab experimental validation confirming high-fold activity improvements in \texttt{Cas12i3}.
    }
    \label{fig:framework}
  \end{center}
\end{figure*}

In this work, we introduce~\model, a two-stage multi-agent framework designed to transition from passive tool execution to active scientific reasoning in Low-$N$ zero-shot protein engineering scenario. By institutionalizing the rigorous review process of LLMs, the system couples the statistical breadth of model ensembles with the biophysical scrutiny of inference-time reasoning. The Rank-Stage utilizes a \textit{Computational Expert} for multi-modal aggregation and a \textit{Virtual Biologist} for context-aware weight calibration. The Reason-Stage instantiates a \textit{Virtual Expert Panel} that employs chain-of-thought reasoning to audit candidates against geometric constraints and local structural confidence. This process operationalizes the tacit intuition of domain experts, filtering statistically confident but biophysically invalid artifacts while recovering high-potential variants that defy consensus ranking.

The effectiveness of~\model~is substantiated through a hierarchical evaluation covering global benchmarks, selection precision, and physical validation. In the Rank-Stage, the framework establishes a new SOTA on \textsc{ProteinGym} with a global Spearman correlation of 0.551, surpassing the previous best of $0.518$. For high-stakes selection scenarios where the experimental budget is limited, the Reason-Stage audit improves the Top-5 Hit Rate by up to $367\%$ and significantly enhances the average Normalized Max Score compared to standard static ensembles. This computational precision translates directly to experimental success; wet-lab validation on the \texttt{Cas12i3} nuclease achieved a $46.7\%$ success rate (14/30 hits), identifying two novel mutants with $4.23$-fold and $5.05$-fold activity improvements, respectively. 
Our contributions are summarized as follows:

\begin{itemize}
    \item \textbf{Paradigm Shift to Scientific Reasoning:} We propose a two-stage agentic architecture that transitions zero-shot protein engineering from passive tool execution to an active, interpretable biological verification process.
    \item \textbf{SOTA Performance and Selection Precision:} We establish a new SOTA for global mutation ranking and demonstrate that agentic auditing provides substantial gains in hit rates and peak fitness discovery under low-resource constraints.
    \item \textbf{Empirical Wet-lab Validation:} We provide proteomic evidence of the framework's utility by successfully discovering novel, high-activity nuclease variants within a strictly constrained experimental budget.
\end{itemize}

\section{Related Work}\label{sec:related_work}
\paragraph{Mutation Prediction Models.}
The landscape of mutation effect prediction has evolved from physics-based simulations to data-driven deep learning. Early approaches, such as \textsc{FoldX} \cite{schymkowitz2005foldx} and \textsc{Rosetta} \cite{das2008rosetta}, relied on thermodynamic stability calculations and structural heuristics. While interpretable, these methods are computationally expensive and sensitive to backbone flexibility. The subsequent wave of co-evolutionary methods, including \textsc{EVE} \cite{frazer2021eve}, \textsc{GEMME} \cite{laine2019gemme}, \textsc{EVmutation} \cite{hopf2017evmutation} and \textsc{DeepSequence} \cite{riesselman2018deepsequence}, leveraged MSAs to capture epistatic constraints, significantly improving zero-shot performance. Recently, PLMs have emerged as the dominant paradigm. Sequence-only models like \textsc{ESM} series \cite{meier2021esm1v,rives2021esm1b,lin2023esm2}, \textsc{PoET} \cite{truong2023poet} and \textsc{ProtTrans} \cite{elnaggar2021prottrans} learn evolutionary patterns from massive unlabeled corpora, enabling rapid scoring of variants via masked language modeling or likelihood estimation. To further enhance predictive accuracy, structure-aware models such as \textsc{ESM-IF1} \cite{hsu2022esm-if1}, \textsc{SaProt} \cite{su2023saprot}, \textsc{ProtSSN} \cite{tan2025protssn} and \textsc{ProSST} \cite{li2024prosst} incorporate 3D geometric priors, while hybrid architectures like \textsc{Tranception} \cite{notin2022tranception,notin2022trancepteve} and \textsc{MSA-Transformer} \cite{rao2021msa}, \textsc{AIDO-Protein} \cite{sun2024aido}, and \textsc{VenusREM} \cite{tan2025venusrem} combine retrieval-based inference with generative modeling. Despite their success on comprehensive benchmarks \cite{notin2024proteingym, zhang2025venusmuthub}, these models operate primarily as static scoring functions. They provide numerical likelihoods without explicit justification, often failing to account for nuanced biophysical trade-offs.

\paragraph{LLM-based Scientific Agents.}
LLMs have transcended their role as text generators to become autonomous agents capable of reasoning, planning, and tool usage \cite{wei2022chain-of-thought, schick2023toolformer}. Systems like \textsc{AI Scientist} \cite{lu2024ai-scientist,yamada2025ai-scientist-v2}, \textsc{AI-Researcher} \cite{tang2025ai-researcher}, \textsc{Agent Laboratory} \cite{schmidgall2025agentlaboratory} are now capable of autonomous literature review, idea generation, code execution, and even full manuscript preparation. In the scientific domain, this agentic paradigm has catalyzed breakthroughs in autonomous discovery, \textsc{ChemCrow} \cite{m2024chemcrow} focuses on solution of reasoning-intensive chemical tasks, \textsc{SciToolAgent} \cite{ding2025scitoolagent} aggregates diverse computational tools to solve multidisciplinary problems, and \textsc{SciAgents} \cite{ghafarollahi2024sciagents} can autonomously accelerate bio-inspired material discovery by integrating ontological knowledge graphs with multi-agent reasoning. Similarly, in biology, agents have been deployed for tasks ranging from protocol generation \cite{o2023bioplanner} to general purpose analyse \cite{huang2025biomni}. \textsc{ProtAgents} \cite{ghafarollahi2024protagents} and \textsc{ProteinCrow} \cite{ponnapati2025proteincrow} established rudimentary agentic pipelines for protein sequence design. \textsc{Virtual Lab} \cite{swanson2025virtuallab} introduced an agent proxy mechanism to guide antibody design, effectively simulating experimental feedback loops. However, an effective multi-agent framework designed to navigate the biophysical constraints and selection precision required for zero-shot protein engineering remains absent.

\section{Method}

\subsection{Problem Formulation}

The task of zero-shot protein engineering is defined as identifying an optimal subset of single-point mutations $\mathcal{X}^*$ within a combinatorial sequence space $\mathcal{V}$ in the absence of experimental labels. We formalize this as a constrained selection problem under a restricted experimental budget $N$.

\paragraph{Formalization of Input Data.}
To integrate statistical inference with agentic reasoning, the input dataset $\mathcal{D}$ is partitioned into two functional components (Figure~\ref{fig:framework}a). (1) The numerical data for PLMs ($\mathcal{D}_{PLM}$) comprises high-dimensional structural and evolutionary data required for PLM inference:
\begin{equation}
    \mathcal{D}_{PLM} = \{ \mathbf{x}_{wt}, \mathbf{G}_{wt}, \mathbf{A}_{wt} \}
\end{equation}
, where $\mathbf{x}_{wt} \in \mathcal{A}^L$ denotes the wild-type sequence of length $L$, $\mathbf{G}_{wt}$ represents the three-dimensional atomic coordinates, and $\mathbf{A}_{wt}$ denotes the MSA. (2) Contextual data for LLMs ($\mathcal{D}_{LLM}$) provides the semantic and biophysical knowledge base utilized by the multi-agent panel: 
\begin{equation}
\mathcal{D}_{LLM} = \{ \text{desc}(\mathcal{M}), \mathcal{C}, \Phi \}
\end{equation}
where $\text{desc}(\mathcal{M})$ signifies natural language descriptions of each model in the ensemble $\mathcal{M}$, $\Phi$ is the formalized engineering objective, and $\mathcal{C}$ is the protein-level (e.g., taxonomy, MSA depth) or residue-level (e.g., $pLDDT$, and relative solvent accessibility) biophysical features.

\paragraph{PLM Modality.}
The framework evaluates the mutation landscape $\mathcal{V}$ to generate statistical scores $\mathcal{S}$ via a modular ensemble $\mathcal{M}$. We formalize $\mathcal{M}$ as a collection of scoring functions across multiple modalities:
\begin{equation}
\mathcal{M} = \bigcup_{t \in \mathcal{T}} \{F_{j,t}\}_{j=1}^{n_t}, \quad \mathcal{T} = \{\text{seq, str, msa}\}
\end{equation}
, where $n_t \ge 1$ permits the integration of multiple heterogeneous PLMs per modality $t$ depending on the data availability in $\mathcal{D}_{PLM}$.

\paragraph{Objective: Expected Maximum Fitness.}
In low-$N$ protein engineering, the objective is to discover the global optimum rather than minimize mean prediction error. To this end, we define the optimization target as maximizing the expected maximum fitness of the selected subset $\mathcal{X}^*$:

\begin{equation}
    \mathcal{X}^* = \arg \max_{\mathcal{X} \subset \mathcal{V}, |\mathcal{X}|=N} \mathbb{E} \left[ \max_{x \in \mathcal{X}} f(x) \right]
\end{equation}

, where $f(x)$ represents the wet-lab fitness of mutant $x$. 

\subsection{Rank-Stage: Context-Aware Statistical Ranking}
The Rank-Stage establishes a robust statistical baseline through the collaborative execution of the \textit{Computational Expert} and the \textit{Virtual Biologist} (Figure~\ref{fig:framework}b and the process details are shown in Appendix Section~\ref{app:sec:rank_stage_prompts}).

\paragraph{Multi-modal Scoring Execution.}
The \textit{Computational Expert} executes the modular ensemble $\mathcal{M}$ by integrating three distinct predictor modalities to capture diverse biological signals. Sequence-based models ($F_{seq}$), such as \textsc{ESM2} and \textsc{ProGen3}, leverage general-purpose representations derived from massive sequence databases, while structure-based models ($F_{str}$), including \textsc{ProSST} and \textsc{ESM-IF1}, assess thermodynamic stability through geometric priors from $\mathbf{G}_{wt}$. Finally, MSA-based models ($F_{msa}$), such as \textsc{GEMME} and \textsc{VenusREM}, are employed to capture functional conservation and epistatic effects from co-evolutionary patterns.

\paragraph{Agentic Dynamic Weighting}
The virtual biologist evaluates the epistemic reliability of input data to calibrate the ensemble weights $\omega$. The final score for a mutant $x$ is formulated as:

\begin{equation}
    \mathcal{S}_{rank}(x) = \sum_{t \in \mathcal{T}} \sum_{j=1}^{n_t} \omega_{j,t}(desc(\mathcal{M}),\mathcal{C}, \Phi) \cdot F_{j,t}(x \mid \mathcal{D}_{PLM})
\end{equation}

This process begins with objective-prior calibration, synthesizing model descriptions ($desc(\mathcal{M})$) to align model inductive biases with the engineering target $\Phi$. To safeguard the ranking against data noise, the agent performs structural calibration—attenuating weights in low-confidence regions (e.g., $pLDDT < 50$) —and evolutionary calibration to marginalize sparse co-evolutionary signals. Such adaptive weighting establishes a high-recall candidate pool for the subsequent reason-stage audit.

\begin{table*}[t]
\centering
\caption{Performance comparison across different protein models on \textsc{ProteinGym} substitution. We report Spearman correlation on various functional tasks. \textbf{Seq.}: Amino acid sequence input; \textbf{Struct.}: Structure input; \textbf{MSA}: MSA input; \textbf{LLM}: LLM-based agent. Model name with * represents the SOTA model on the leaderboard\footnote{\url{https://proteingym.org/benchmarks}}.}
\label{tab:proteingym}
\resizebox{\textwidth}{!}{%
\begin{tabular}{@{}lcccccccccc@{}}
    \toprule
    \multirow{2}{*}{\textbf{Model}} & \multicolumn{4}{c}{\textbf{Inputs}} & \multirow{2}{*}{\textbf{Avg. Spearman}} & \multicolumn{5}{c}{\textbf{Spearman by Function}} \\ \cmidrule(lr){2-5} \cmidrule(l){7-11} 
     & \textbf{Seq.} & \textbf{Struct.} & \textbf{MSA} & \textbf{LLM} &  & \textbf{Activity} & \textbf{Binding} & \textbf{Expression} & \textbf{Org. Fitness} & \textbf{Stability} \\ 
     \midrule
    \textsc{ProGen3-3B} & \y & \n & \n & \n & 0.392 & 0.410 & 0.287 & 0.428 & 0.400 & 0.438 \\
    \textsc{ESM-C} & \y & \n & \n & \n & 0.406 & 0.423 & 0.315 & 0.408 & 0.360 & 0.526 \\
    \textsc{ESM2-650M} & \y & \n & \n & \n & 0.414 & 0.425 & 0.337 & 0.415 & 0.368 & 0.523 \\ 
    \midrule
    \textsc{ProtSSN-Ensemble} & \y & \y & \n & \n & 0.449 & 0.466 & 0.366 & 0.449 & 0.396 & 0.568 \\
    \textsc{SaProt-AF650M} & \y & \y & \n & \n & 0.457 & 0.458 & 0.378 & 0.488 & 0.366 & 0.592 \\
    \textsc{ProSST-2048} & \y & \y & \n & \n & 0.507 & 0.476 & 0.455 & 0.530 & 0.431 & 0.653 \\ 
    \midrule
    \textsc{MSA-Transformer} & \y & \n & \y & \n & 0.432 & 0.473 & 0.329 & 0.446 & 0.419 & 0.492 \\
    \textsc{GEMME} & \y & \n & \y & \n & 0.455 & 0.482 & 0.383 & 0.438 & 0.452 & 0.519 \\
    \textsc{TranceptEVE-L} & \y & \n & \y & \n & 0.456 & 0.487 & 0.376 & 0.457 & 0.459 & 0.500 \\ 
    \midrule
    \textsc{S3F-MSA} & \y & \y & \y & \n & 0.496 & 0.502 & 0.440 & 0.479 & 0.477 & 0.581 \\
    \textsc{AIDO-Protein-RAG}* & \y & \y & \y & \n & 0.518 & 0.517 & 0.426 & 0.522 & 0.491 & 0.635 \\
    \textsc{VenusREM}* & \y & \y & \y & \n & 0.518 & 0.495 & 0.454 & 0.533 & 0.459 & 0.650 \\ 
    \midrule
    \textsc{Deepseek-Reasoner} & \y & \n & \n & \y & 0.159 & 0.149 & 0.149 & 0.158 & 0.154 & 0.187 \\
    \midrule
    \model-Ensemble & \y & \y & \y & \n & 0.542 & 0.533 & 0.478 & 0.556 & 0.485 & \textbf{0.661} \\
    \textbf{\model-Rank} & \y & \y & \y & \y & \textbf{0.551} & \textbf{0.539} & \textbf{0.497} & \textbf{0.558} & \textbf{0.505} & 0.658 \\ 
    \bottomrule
    \end{tabular}%
}
\end{table*}

\begin{figure*}[ht]
  \vskip 0.2in
  \begin{center}
    \centerline{\includegraphics[width=\textwidth]{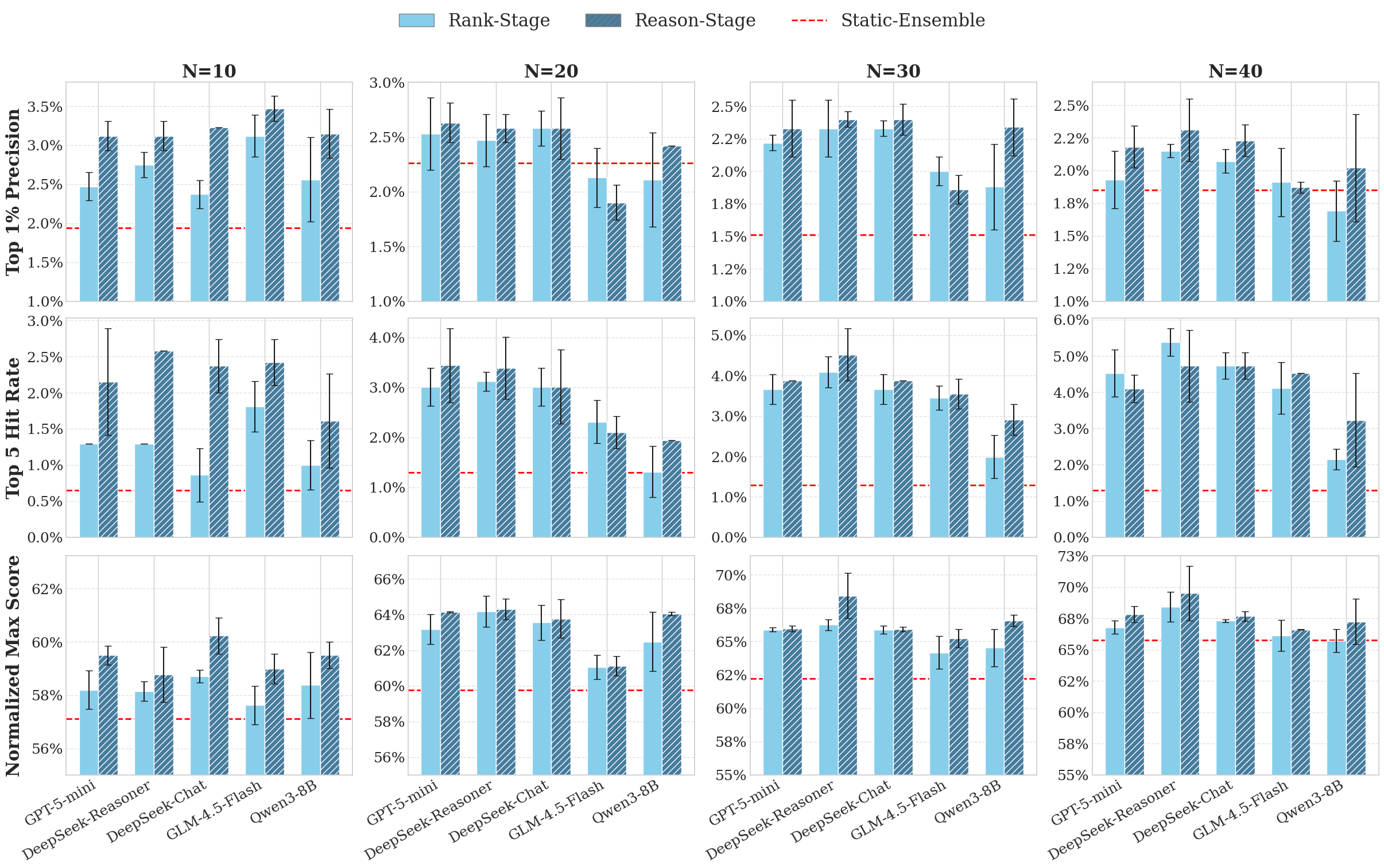}}
    \caption{
      Comparison of Rank-Stage (solid bars) versus Reason-Stage (hashed bars) across budgets $N \in \{10, \dots, 40\}$. Panels display Top 1\% Precision (top), Top 5 Hit Rate (middle), and Normalized Max Score (bottom). The red dashed line marks the Static-Ensemble. 
    }
    \label{fig:stage_comparison}
  \end{center}
\end{figure*}

\subsection{Reason-Stage: Physics-Informed Agentic Reasoning}
The Reason-Stage transitions from statistical aggregation to residue-specific verification via a multi-agent \textit{Virtual Expert Panel} (Figure~\ref{fig:framework}c with details in Appendix Section~\ref{app:sec:reason_stage_prompts}).

\paragraph{Virtual Expert Panel.}
Three specialized agents: (1) the \textit{Statistical Auditor}, which manages the comprehensive candidate pool $\mathcal{P}$, provides analytical insights into ensemble ranking inconsistencies, and enforces positional diversity to maximize functional coverage; (2) the \textit{Structural Biologist}, which evaluates residue-level structural integrity and geometric stability via $pLDDT_i$, executing a conditional trust policy that prioritizes evolutionary consensus in low-confidence structural regions ($pLDDT_i < 50$); and (3) the \textit{Experimental Expert}, which ensures wet-lab feasibility by assessing developability profiles, utilizing features such as $RSA$ and net charge to flag potential expression risks while applying a biophysical exclusion set $\mathcal{E}$ for reactive or disruptive residues.

\paragraph{Candidate Pool Construction.}
To ensure high recall, the \textit{Statistical Auditor} first constructs a candidate pool $\mathcal{P} \subset \mathcal{V}$. Let $n_{total} = |\mathcal{M}|$ be the total number of unique scoring models within the ensemble. The auditor aggregates the top $K=200$ variants from the calibrated ensemble ranking $\mathcal{S}_{rank}$ and the top $K$ variants from each of the $n_{total}$ individual models. The search space is thus bounded by:
\begin{equation}
    |\mathcal{P}| \le (n_{total} + 1) \times 200
\end{equation}
This formulation allows the system to recover high-potential variants that may be underestimated by the ensemble average but are highly ranked by specialized component models.

\paragraph{Iterative Audit and Replacement.}
Candidates within $\mathcal{P}$ are evaluated in descending order of their statistical scores $\mathcal{S}_{rank}$ against a joint boolean audit function:
\begin{equation}
    \mathcal{A}(x, \mathcal{C}_i) = \text{Val}_{stat}(x) \land \text{Val}_{str}(x, \mathcal{C}_i) \land \text{Val}_{exp}(x, \mathcal{C}_i)
\end{equation}
, where $\text{Val}_{stat}$, $\text{Val}_{str}$, and $\text{Val}_{exp}$ represent the consensus of the statistical, structural, and experimental auditors. The selection logic follows a deterministic sequence where a candidate $x$ is added to the selection set $\mathcal{X}^*$ if it satisfies all agentic constraints ($\mathcal{A}=1$), while a detected violation ($\mathcal{A}=0$) triggers a replacement search, leading to the candidate's rejection and an immediate audit of the next-best variant in the priority sequence. This reasoning-based optimization terminates once $|\mathcal{X}^*|= N$, where $N$ is the pre-defined experimental budget, ensuring the final subset maximizes expected maximum fitness while adhering to rigorous biophysical requirements for developability.

\section{Experiments}
\subsection{Experimental Setup}

\paragraph{Configuration.}
We construct a heterogeneous expert ensemble for Rank-Stage using $6$ SOTA models: \textsc{VenusREM} , \textsc{ProSST-2048}, \textsc{SaProt-AF650M}, \textsc{ProtSSN-Ensemble}, \textsc{GEMME}, and \textsc{ESM-IF1}. These models were selected for their complementary coverage of biological modalities (sequence, structure, and MSA) and have been proven reliable in wet-lab validations \cite{zhou2024fsfp,tan2025venusrem,su2025saprothub,johnson2025enzyme_wet_lab,Zheng2026protssn_wet_lab}. For the reasoning backbone, we employ \textsc{DeepSeek-Resoner} in Rank-Stage during the initial ranking. In Reason-Stage, we benchmark a diverse array of LLMs with varying parameter scales and architectures, including \textsc{GPT-5-mini}, \textsc{DeepSeek} series (Chat/Reasoner) \cite{guo2025deepseek}, \textsc{GLM} \cite{du2021glm}, and \textsc{Qwen} \cite{yang2025qwen3}. Details are provided in Appendix Section~\ref{app:sec:exp_detail}.

\paragraph{Datasets.}
For the global ranking evaluation, we utilize the \textsc{ProteinGym} substitutions benchmark \cite{notin2024proteingym}, which comprises $217$ high-throughput Deep Mutational Scanning (DMS) datasets covering over $2$ million mutations. This represents the most comprehensive zero-shot benchmark in the field.
For the selection capability evaluation, verifying the Top Hit Rate requires complete ground truth to avoid false negatives caused by missing data. Therefore, we curated a high-fidelity subset named \textsc{ProteinGym-DMS99}, consisting of 31 DMS datasets where single-point mutation coverage exceeds 99\%. Static information at Appendix Table~\ref{app:tab:proteingym_stats_general} and Table~\ref{app:tab:proteingym_stats_distribution}.

\paragraph{Baseline Models.}
In the Rank-Stage, we compare \model~against a comprehensive set of baselines spanning four paradigms:
(1) Sequence-only models; (2) Structure-conditioned models; (3) MSA-based models; (4) Hybrid models.
In the Reason-Stage, the primary comparison focuses on the performance differential between Rank-Stage, Reason-Stage, and varying LLM backends. Additionally, to demonstrate the efficacy of agentic workflow, we compare against \model-Ensemble (arithmetic mean of the selected expert models) and a direct LLM zero-shot approach. Details can be found in Appendix Section~\ref{app:sec:exp_detail}.

\paragraph{Evaluation Metrics.}
Global performance is measured using Spearman’s rank correlation coefficient on the full \textsc{ProteinGym} dataset. To capture the nuances of real-world wet-lab constraints, we evaluate local selection on the high-fidelity \textsc{ProteinGym-DMS99} benchmark under fixed budgets of $N \in \{10, 20, 30, 40\}$. These selection-centric metrics include: (1) normalized max score, which identifies the highest DMS score within the budget $N$, min-max normalized to $[0, 1]$ per protein; (2) precision (top $X$\%), representing the proportion of candidates within the budget that rank among the top $X$\% of the fitness landscape for $X \in \{1, 5, 10\}$; and (3) hit rate (top $K$), quantifying the recovery of the absolute top $K$ functional variants within the allocated area for $K \in \{5, 10, 30\}$.

\subsection{Global Ranking Performance}
We first assess global and functional ranking performance on the comprehensive \textsc{ProteinGym} benchmark (Table~\ref{tab:proteingym} and Appendix Table~\ref{app:tab:add_proteingym}). LLM token computational costs are detailed in Appendix Tables~\ref{app:tab:token_usage} and~\ref{app:tab:dataset_stats}, and different LLM backbones of Rank-Stage evaluations can be found at Appendix Table~\ref{app:tab:llm_backbone_comparison}.

\paragraph{\model~Achieves SOTA.} 
Our results reveal a clear hierarchy (Table~\ref{tab:proteingym} and Appendix Table~\ref{app:tab:add_proteingym}): predictive precision scales with modal integration. Sequence-only models like \textsc{ESM2} achieve $\rho=0.414$, sequence-structure models like \textsc{ProSST} reach $0.507$, and sequence-structure with MSA retrieval-augmented hybrids like \textsc{VenusREM} attain $0.518$ on overall \textsc{ProteinGym}. \model-Rank, which represents the first strategic integration of the LLM modality alongside traditional biological channels, culminates this trajectory, establishing a new SOTA of $0.551$. In stark contrast, standalone \textsc{DeepSeek-Reasoner} achieves only $0.159$, demonstrating that the scientific utility of LLMs is maximized when acting as reasoning orchestrators for expert models rather than as independent predictors.

\paragraph{Robustness of Agentic Improvement.}
A notable finding emerges from our backbone analysis: while agentic orchestration consistently outperforms the static baseline ($\rho=0.542$), the specific LLM backbone exerts minimal influence, with Spearman correlations tightly clustered between $0.543$ and $0.551$ across all models (Appendix Table~\ref{app:tab:llm_backbone_comparison}). This robustness is confirmed by near-identical performance across taxonomic categories and independent runs (Appendix Table~\ref{app:tab:llm_multi_run}). These results indicate that the Rank-Stage functions primarily as a statistical aggregation task. Consequently, performance saturates even with small LLMs; it relies on weight calibration rather than reasoning.

\begin{figure}[t]
  \begin{center}
    \centerline{\includegraphics[width=\linewidth]{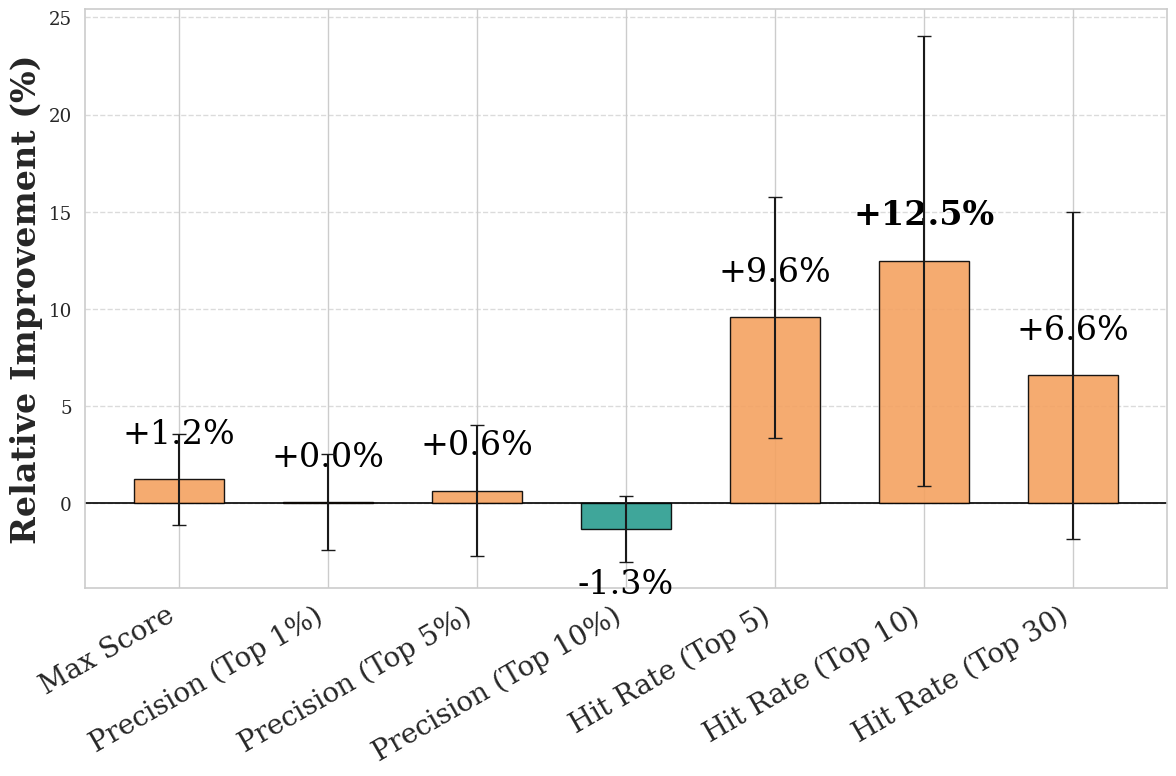}}
    \caption{
      Relative improvement on reasoning capability of \textsc{DeepSeek-Reasoner} vs. \textsc{DeepSeek-Chat}, averaged across different budgets ($N \in \{10, \dots, 40\}$). Error bars denote the standard deviation. 
    }
    \label{fig:reasoning_gap}
  \end{center}
\end{figure}

\subsection{Efficacy of Agentic Auditing in Selection}

Global correlation measures ranking monotonicity across the entire fitness landscape, but practical protein engineering demands precision at the top of the ranked list under constrained budgets. We therefore shift focus to selection metrics on the high-fidelity \textsc{ProteinGym-DMS99} subset. To assess the relative contribution of the two-stage pipeline under realistic Low-$N$ protein engineering scenarios, we evaluate the model performance across $N \in \{10, 20, 30, 40\}$ (Figure~\ref{fig:stage_comparison}, Figure~\ref{fig:reasoning_gap} and Appendix Figure~\ref{app:fig:max_score}). Numerical details are shown in Appendix Table~\ref{app:tab:rar_budget_full}.

\paragraph{Consistent Uplift via Auditing.}
Across all evaluated configurations (Figure~\ref{fig:stage_comparison} and Appendix Table~\ref{app:tab:rar_budget_full}), Reason-Stage consistently outperforms the Rank-Stage statistical baseline. Notably, at a budget of $N=40$, the integration of the agentic audit mechanism enhances the Top 5 Hit Rate by an average of $15-20\%$. This confirms that the multi-agent panel effectively filters biophysically inviable candidates—statistical artifacts that standard PLMs often categorize with high confidence but violate fundamental geometric or structural constraints.

\paragraph{Correlation-Precision Gap.}

A critical insight emerges from comparing global and local metrics. While the static \model-Ensemble baseline achieves a competitive global Spearman correlation ($\rho=0.542$), its practical utility collapses in high-stakes selection scenarios, but most of the Top 1\% Precisions, Top 5 Hit Rates, and Normalized Max Scores lag significantly behind the Reason-Stage Figure~\ref{fig:stage_comparison}. Illustrated by the Top 5 Hit Rate, where the Reason-Stage achieves a massive 367\% improvement over the Static-Ensemble. Such divergence confirms that global correlation masks high-confidence false positives, which are effectively eliminated only through the biophysical discrimination of the agentic audit.

\paragraph{Reasoning Capacity Dictates Audit Reliability.}
The efficacy of the Reason-Stage is strictly governed by the capacity of the LLM backbone (Figure~\ref{fig:stage_comparison}). Advanced models like \textsc{DeepSeek-Reasoner} and \textsc{GPT-5-mini} demonstrate superior robustness, maintaining high Top 1\% Precision even under low budget ($N=10$), whereas smaller architectures (e.g., \textsc{Qwen3-8B}) exhibit significant stochasticity. This impact of inference-time compute is further quantified by the performance differential between \textsc{DeepSeek-Reasoner} and its standard counterpart \textsc{DeepSeek-Chat} (Figure~\ref{fig:reasoning_gap}). The resulting gap confirms that advanced chain-of-thought capability is not merely an enhancement but a functional prerequisite for the effective self-correction required to purify candidate pools.

\begin{table}[t]
\centering
\caption{Performance comparison (\%) of mutation selection based solely on LLMs under budget $N=30$. The format is reported as $Mean_{(Std)}$. \textbf{Seq.}: Amino acid sequence; \textbf{Str.}: Protein structure; \textbf{Ann.}: Protein annotation information.}
\label{tab:llm_zero-shot}
\resizebox{\linewidth}{!}{%
\begin{tabular}{@{}llccc@{}}
    \toprule
    \textbf{LLM} & \textbf{Inputs} & \textbf{Max} & \textbf{Top 1\%} & \textbf{Top 5\%} \\ \midrule
    Random & - & $58.41_{(3.40)}$ & $0.85_{(0.28)}$ & $4.41_{(0.67)}$ \\ 
    \midrule
    \multirow{3}{*}{\textsc{GPT-5-mini}} & Seq & $58.17_{(1.90)}$ & $0.85_{(0.28)}$ & $6.37_{(0.67)}$ \\
     & Seq.+Ann. & $57.72_{(0.90)}$ & $0.71_{(0.06)}$ & $6.74_{(0.28)}$ \\
     & Seq.+Ann.+Str. & $59.00_{(1.23)}$ & \textbf{$1.15_{(0.45)}$} & \textbf{$7.37_{(0.74)}$}  \\ 
     \midrule
    \multirow{3}{*}{\textsc{DeepSeek-Reasoner}} & Seq & $55.52_{(2.87)}$ & $0.70_{(0.23)}$ & $5.78_{(0.62)}$ \\
     & Seq.+Ann. & \textbf{$60.12_{(1.77)}$} & $1.15_{(0.55)}$ & $7.11_{(1.26)}$ \\
     & Seq.+Ann.+Str. & $57.61_{(1.55)}$ & $0.78_{(0.51)}$ & $6.92_{(0.57)}$ \\ 
     \bottomrule
\end{tabular}%
}
\end{table}

\begin{figure*}[t]
  \begin{center}
    \centerline{\includegraphics[width=1\textwidth]{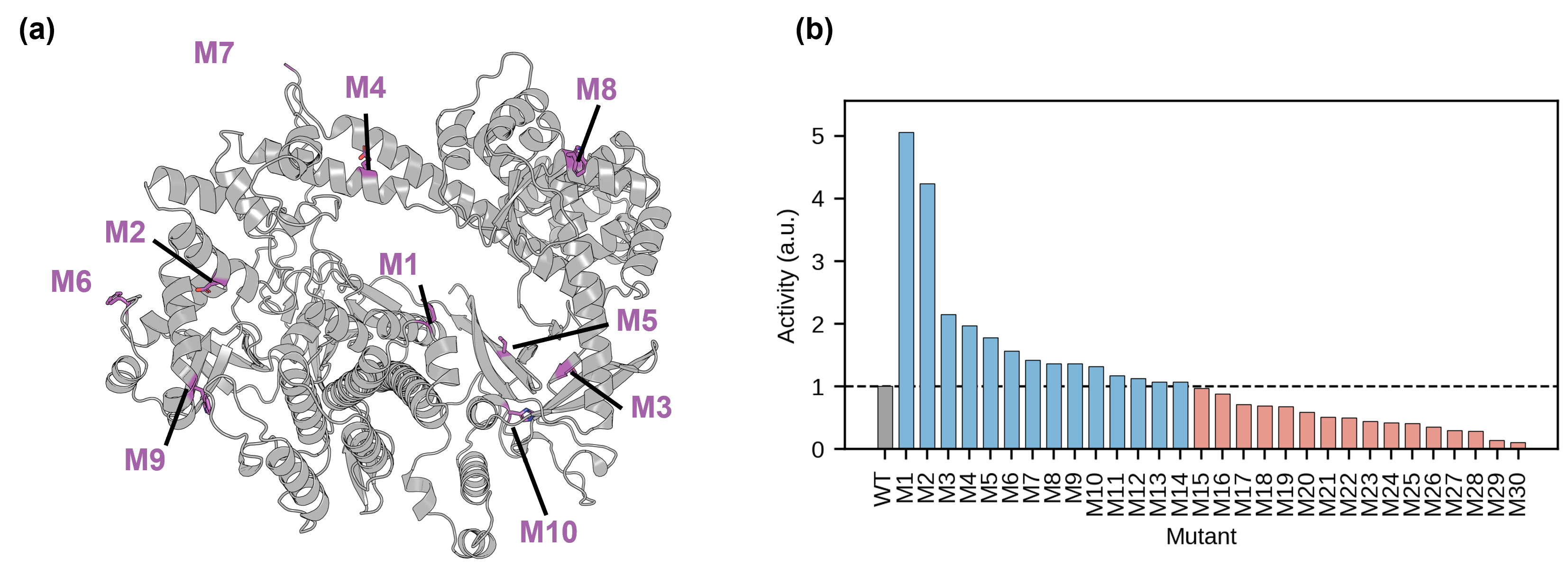}}
    \caption{
      Wet-lab validation on \texttt{Cas12i3}. \textbf{(a)} The 3D distribution of the identified hits. Beneficial mutations are dispersed across diverse structural domains, verifying the system's global search capability. \textbf{(b)} Activity landscape of the 30 candidates relative to Wild Type (WT=1.0). \model~achieves a $46.7\%$ positive rate ($14/30$), with the top variant~\ma~($5.05$-fold).
    }
    \label{fig:wet-lab}
  \end{center}
\end{figure*}

\subsection{Study of LLMs Capabilities}
To understand the necessity of our agentic framework, we conducted a granular ablation study on standalone LLMs in Table~\ref{tab:llm_zero-shot}, Appendix Table~\ref{app:tab:llm_zero_shot_full} and~\ref{app:tab:llm_few_shot}.

\paragraph{Effect of Input Modalities.}
Ablation studies on LLMs' zero-shot selection reveal a distinct dichotomy in multi-modal processing. Textual annotations consistently enhance precision—boosting \textsc{GPT-5-mini}'s Top 10\% Precision from $13.37\%$ to $14.11\%$ (Appendix Table~\ref{app:tab:llm_zero_shot_full})—indicating effective semantic grounding. Conversely, raw structural coordinates act as geometric noise, degrading \textsc{DeepSeek-Reasoner}'s Max Score from $60.12$ to $57.61$ (Table~\ref{tab:llm_zero-shot}). This confirms that current LLMs function primarily as semantic processors rather than geometric encoders; without agentic weighting, linearized 3D data acts as entropic interference that disrupts the reasoning chain.

\paragraph{Absence of LLMs Data Leakage.}
A persistent concern in LLM benchmarking is potential data leakage from pre-training corpora. Both zero-shot and few-shot analyses reveal linear performance scaling ($\rho \in [0.3, 0.5]$) rather than the step-function gains characteristic of memorization (Appendix Tables~\ref{app:tab:llm_zero_shot_full},~\ref{app:tab:llm_few_shot}).  
Critically, standalone LLMs perform below random at $N=40$ ($59.05$--$59.15$ vs. $61.09$), confirming the absence of data leakage.  A functional divergence also emerges: \textsc{GPT-5-mini} excels in in-context ranking ($\rho=0.521$) while \textsc{DeepSeek-Reasoner} plateaus, validating our design of delegating statistical ranking to generalists while reserving reasoners for structural auditing.

\subsection{Wet-lab Validation}
To validate~\model~in a realistic Low-$N$ scenario, we applied it to enhance \texttt{Cas12i3} nuclease activity under a budget of $N=30$ (protein function description, web-lab protocol, and agent decision details are in Appendix Section~\ref{app:sec:agent_cas12i3}). Structurally, these hits are dispersed across distinct domains rather than clustering in a single region (Figure \ref{fig:wet-lab}a), confirming the effectiveness of our diversity constraints. Furthermore, all validated variants occupy geometrically feasible rotamers consistent with local backbone stability, validating the precision of the \textit{Statistical Auditor}. 

The experimental results, visualized in Figure \ref{fig:wet-lab}, substantiate the efficacy of our framework. Remarkably, $14$ out of the $30$ selected variants exhibited activity superior to the wild type (WT=1.0), translating to a $46.7\%$ positive rate. Among the confirmed hits, we identified two novel super-mutants, \ma~and~\mb, which achieved 5.05-fold and 4.23-fold activity improvements respectively. This combination of high functional diversity and structural viability not only proves the framework's zero-shot efficiency but also establishes an optimal, robust starting library for subsequent combinatorial directed evolution.

\section{Conclusion}
Protein mutation prediction stands as a cornerstone of industrial biotechnology, driving substantial economic value through the design of high-performance biocatalysts. This domain mirrors the broader evolution of scientific discovery, transitioning from first-principles physical calculations to data-driven deep learning, and now towards autonomous agentic systems. Despite this progress, we expose a critical disconnect in current ensemble pipelines: they prioritize statistical correlation over biological reality, frequently yielding high-confidence hallucinations that fail in experimental validation. We address this challenge by introducing a rationally designed two-stage multi-agent framework (\model) that bridges the epistemic gap between passive tool execution and active scientific reasoning. By institutionalizing biophysical constraints within the inference process, our approach significantly enhances the hit rate of wet-lab experiments, proving that architectural rationality is the key to unlocking the next frontier of protein design.

Beyond performance, a key design philosophy of \model~ is its plug-and-play modularity. Recognizing that domain experts often utilize proprietary predictors or specific LLMs, our framework is designed to be backend-agnostic, allowing users to seamlessly define their trusted expert models for the Rank-Stage ensemble or swap the LLM backbone. Looking ahead, we aim to extend the ``Rank-and-Reason" paradigm to combinatorial design, where epistatic complexity demands deeper reasoning. We also plan to incorporate active learning loops to refine the \textit{Virtual Biologist}'s judgment using wet-lab feedback, and explore multi-modal LLMs that can directly perceive 3D protein structures to further enhance agentic auditing capabilities.




\section*{Impact Statement}
Our work on the ``Rank-and-Reason" paradigm can be used in developing potent biocatalysts and therapeutics and accelerate the research process of drug discovery by bridging the gap between prediction and experimental reality. Our framework may be adapted to other scenarios of active agentic auditing, such as small molecule design, material design, and chemical synthesis. It is also needed to ensure the responsible use of our method and refrain from using it for harmful purposes.

\nocite{langley00}

\bibliography{1reference}
\bibliographystyle{icml2026}

\newpage
\appendix
\onecolumn

\section{Experimental Details}\label{app:sec:exp_detail}
\subsection{LLM Detail}

\begin{table}[htbp]
\centering
\caption{Details of Large Language Models: versions, thinking capability, and implementation sources.}
\label{app:tab:llm_details}
\resizebox{\textwidth}{!}{%
    \begin{tabular}{@{}llcl@{}}
    \toprule
    \textbf{LLM} & \textbf{Version} & \textbf{Thinking} & \textbf{Implementation} \\ \midrule
    \textsc{DeepSeek-Reasoner} & v3.2 & $\y$ & \url{https://api-docs.deepseek.com/} \\
    \textsc{DeepSeek-Chat} & v3.2 & $\n$ & \url{https://api-docs.deepseek.com/} \\
    \textsc{GPT-5-mini} & - & $\y$ & \url{https://platform.openai.com/docs/models/gpt-5-mini} \\
    \textsc{Qwen3-8B} & - & $\n$ & \url{https://github.com/QwenLM/Qwen3} \\
    \textsc{GLM-4.5-Flash} & - & $\n$ & \url{https://docs.bigmodel.cn/cn/guide/models/free/glm-4.5-flash} \\ \bottomrule
    \end{tabular}%
}
\end{table}

Table~\ref{app:tab:llm_details} details the specifications of the large language models utilized in this study, categorizing them by version, reasoning capability (Thinking), and official implementation source.

\subsection{Expert Model Detail}

\begin{table}[htbp]
\centering
\caption{Summary of baseline models, input modalities, model types, and official code repositories.}
\label{app:tab:plm_baseline_implementation}
\resizebox{\textwidth}{!}{%
    \begin{tabular}{@{}llll@{}}
    \toprule
    \textbf{Model} & \textbf{Input Modalities} & \textbf{Type} & \textbf{Implementation} \\ \midrule
    \textsc{ProGen3-3B} & - & Decoder & \url{https://github.com/Profluent-AI/progen3} \\
    \textsc{ESM-C} & - & Encoder& \url{https://github.com/evolutionaryscale/esm} \\
    \textsc{ESM2-650M} & - & Encoder & \url{https://github.com/facebookresearch/esm} \\
    \textsc{ProtSSN-Ensemble} & Structure & Encoder & \url{https://github.com/ai4protein/ProtSSN} \\
    \textsc{SaProt-AF650M} & Structure & Encoder & \url{https://github.com/westlake-repl/SaProt} \\
    \textsc{ProSST-2048} & Structure & Encoder & \url{https://github.com/ai4protein/ProSST} \\
    \textsc{MSA-Transformer} & MSA & Encoder & \url{https://github.com/facebookresearch/esm} \\
    \textsc{GEMME} & MSA & Statistical & \url{https://www.lcqb.upmc.fr/GEMME/} \\
    \textsc{TranceptEVE-L} & MSA & Decoder & \url{https://github.com/OATML-Markslab/Tranception} \\
    \textsc{S3F-MSA} & Structure \& MSA & Encoder & \url{https://github.com/DeepGraphLearning/S3F} \\
    \textsc{AIDO-Protein-RAG} & Structure \& MSA & Decoder & \url{https://huggingface.co/genbio-ai/AIDO.Protein-RAG-16B} \\
    \textsc{VenusREM} & Structure \& MSA & Encoder & \url{https://github.com/ai4protein/VenusREM} \\ \bottomrule
    \end{tabular}%
}
\end{table}

Table~\ref{app:tab:plm_baseline_implementation} provides a comprehensive summary of the baseline models employed in this study, detailing their respective input modalities, architectural types, and official implementation sources.

\paragraph{Sequence-only models:}
\begin{itemize}
    \item \textsc{ProGen3} \cite{bhatnagar2025progen3}: An autoregressive decoder-only PLM pretrained on the Profluent Protein Atlas, a curated corpus of billions of natural protein sequences, using a sparse mixture-of-experts Transformer architecture for large-scale protein generation and design.
    \item \textsc{ESM-C} \cite{hayes2025esm3}: A transformer encoder pretrained with masked language modeling on large-scale public protein sequence databases (e.g., UniRef) to learn conditional protein representations.
    \item \textsc{ESM2} \cite{lin2023esm2}: A masked language Transformer encoder trained on UniRef-scale protein sequence datasets to learn general-purpose representations capturing structural and functional information.
\end{itemize}

\paragraph{Structure-conditioned models:}
\begin{itemize}
    \item \textsc{ProtSSN} \cite{tan2025protssn}: A structure-aware protein representation model that incorporates 3D structural information into a transformer-based encoder, pretrained on paired protein sequence and structure data.
    \item \textsc{SaProt} \cite{su2023saprot}: A structure-conditioned PLM pretrained on protein sequences augmented with structures encoded by \textsc{Foldseek} \cite{van2024foldseek} to enhance structure-function representation learning.
    \item \textsc{ProSST} \cite{li2024prosst}: A transformer-based model that jointly models protein sequences and structural features, leveraging disentangled attention to improve sequence–structure understanding.
\end{itemize}

\paragraph{MSA-based models:} 
\begin{itemize}
    \item \textsc{MSA Transformer} \cite{rao2021msa}: A masked language model operating directly on multiple sequence alignments, pretrained on millions of MSAs to capture evolutionary and co-variation signals through row and column attention.
    \item \textsc{GEMME} \cite{laine2019gemme}: A statistical evolutionary model that predicts mutational effects by explicitly modeling global epistasis from homologous sequence alignments without deep neural network training.
    \item \textsc{TranceptEVE} \cite{notin2022trancepteve}: A hybrid framework combining a family-agnostic autoregressive PLM with a family-specific variational model derived from MSAs for improved fitness prediction.
\end{itemize}

\paragraph{Hybrid models:}
\begin{itemize}
    \item \textsc{S3F-MSA} \cite{zhang2024s3f}: A multi-view protein representation model that fuses sequence, structure, and MSA information using a transformer-based architecture to capture complementary evolutionary and geometric signals.
    \item \textsc{AIDO-Protein-RAG} \cite{sun2024aido}: A large-scale generative protein foundation model that integrates sequence and evolutionary information via a decoder architecture with retrieval-augmented generation.
    \item \textsc{VenusREM} \cite{tan2025venusrem}: A multi-modal protein representation model pretrained on sequence, structure, and enhanced with MSA inputs retrieval mechanism for mutation prediction.
\end{itemize}

\section{Dataset Description}

\newcommand{\res}[2]{#1_{(#2)}}
\begin{table}[ht]
\centering
\caption{General statistics of ProteinGym and ProteinGym-DMS99, including sequence length, structural confidence (pLDDT), and evolutionary depth ($N_{\text{eff}}$).}
\label{app:tab:proteingym_stats_general}
\resizebox{0.7\textwidth}{!}{%
\begin{tabular}{@{}lccccc@{}}
    \toprule
    \textbf{Dataset} & \textbf{\# Protein} & \textbf{Avg. Len} & \textbf{Avg. pLDDT} & \textbf{Avg. $N_{\text{eff}}$} & \textbf{Avg. $N_{\text{eff}}/L$} \\ 
    \midrule
    \textsc{ProteinGym} & 217 & 397.14 & 86.94 & 24330.65 & 265.76 \\
    \textsc{ProteinGym-DMS99} & 31 & 304.97 & 87.88 & 12918.24 & 151.74 \\ 
    \bottomrule
\end{tabular}%
}
\end{table}

\begin{table}[ht]
\centering
\caption{Distribution of datasets across MSA Depth levels (High, Medium, Low) and Functional categories.}
\label{app:tab:proteingym_stats_distribution}
\resizebox{0.8\textwidth}{!}{%
\begin{tabular}{@{}lcccccccc@{}}
    \toprule
    \multirow{2}{*}{\textbf{Dataset}} & \multicolumn{3}{c}{\textbf{MSA Depth}} & \multicolumn{5}{c}{\textbf{Function}} \\ 
    \cmidrule(lr){2-4} \cmidrule(l){5-9}
     & \textbf{High} & \textbf{Medium} & \textbf{Low} & \textbf{Activity} & \textbf{Binding} & \textbf{Expression} & \textbf{Org. Fitness} & \textbf{Stability} \\ 
     \midrule
    \textsc{ProteinGym} & 72 & 109 & 36 & 43 & 13 & 18 & 77 & 66 \\
    \textsc{ProteinGym-DMS99} & 7 & 17 & 7 & 4 & 0 & 1 & 15 & 11 \\ 
    \bottomrule
\end{tabular}%
}
\end{table}

\begin{table}[htbp]
\centering
\begin{minipage}{0.58\linewidth}
\centering
\caption{Token usage statistics.}
\label{app:tab:token_usage}
\begin{tabular}{lccc}
    \hline
    \textbf{Type} & \textbf{Total Tokens} & \textbf{Avg / Protein} & \textbf{Avg / Mutation} \\
    \hline
    Input  & 51,172,532  & 235,818.12 & 20.75 \\
    Output & 197,907,475 & 912,016.01 & 80.26 \\
    \hline
\end{tabular}
\end{minipage}
\hfill
\begin{minipage}{0.38\linewidth}
\centering
\caption{Dataset statistics.}
\label{app:tab:dataset_stats}
\begin{tabular}{lc}
    \hline
    \textbf{Metric} & \textbf{Value} \\
    \hline
    \# Proteins   & 217 \\
    \# Mutations  & 2,465,767 \\
    I/O Ratio    & 0.2586 \\
    \hline
\end{tabular}
\end{minipage}
\end{table}

In this section, we provide a comprehensive statistical breakdown of the datasets utilized for evaluation, specifically the full \textsc{ProteinGym} benchmark ($N=217$) and its subset \textsc{ProteinGym-DMS99} ($N=31$). As summarized in Table~\ref{app:tab:proteingym_stats_general}, the proteins maintain high structural confidence (Avg. pLDDT $> 86$) while spanning a broad range of sequence lengths and evolutionary depths ($N_{\text{eff}}$). To ensure a robust assessment of generalization capabilities, the datasets encompass diverse functional categories—ranging from organism fitness to protein stability—and varying MSA depth profiles (Table~\ref{app:tab:proteingym_stats_distribution}). Furthermore, Table~\ref{app:tab:token_usage} and Table~\ref{app:tab:dataset_stats} illustrate the extensive computational scale of our agentic framework, which processed over 2.4 million mutations and generated approximately 198 million output tokens to facilitate deep reasoning.

\section{System Prompts of Rank-Stage Agents}\label{app:sec:rank_stage_prompts}

This section details the system prompts utilized in the Rank-Stage. The framework orchestrates two specialized agents sequentially: the \textit{Virtual Biologist} and the \textit{Computational Expert}.

\subsection{Agent 1: Virtual Biologist (Bio-Data Profiling)}

The \textit{Virtual Biologist} agent analyzes raw metadata and structural quality indicators to generate a qualitative biological profile. It provides the necessary biophysical context (the "Knowledge") that guides the subsequent weighting strategy.

\begin{promptbox}{System Prompt: Virtual Biologist}
Updated upon paper acceptance.
\end{promptbox}

\paragraph{Parameter Explanations:}
\begin{itemize}
    \item \placeholder{molecule\_name}, \placeholder{taxon}: Basic identification of the target protein.
    \item \placeholder{MSA\_N\_eff}: The effective number of sequences in the MSA, indicating evolutionary richness.
    \item \placeholder{mean\_plddt}, \placeholder{high\_conf\_ratio}: Statistical summaries derived from AlphaFold2/PDB B-factors, quantifying 3D structural reliability.
\end{itemize}

\subsection{Agent 2: Computational Expert (Strategy Generation)}

The \textit{Computational Expert} agent acts as the strategic executor. It translates the qualitative biological profile into a concrete, executable Python weighting strategy, dynamically adjusting the ensemble based on the reliability of available data modalities.

\begin{promptbox}{System Prompt: Computational Expert}
Updated upon paper acceptance.
\end{promptbox}

\paragraph{Parameter Explanations:}
\begin{itemize}
    \item \placeholder{model\_list\_str}: A dynamically generated list of available expert model columns (e.g., "- \texttt{VenusREM\_score}").
    \item \placeholder{knowledge}: The bio-data profile generated by the \textit{Virtual Biologist} in the previous step, which serves as the conditional context for weight calibration.
\end{itemize}

\section{System Prompts of Reason-Stage Agents}
\label{app:sec:reason_stage_prompts}

In the Reason-Stage, the framework instantiates a \textit{Senior Expert Review Panel} comprising three distinct specialist roles: a \textit{Statistical Auditor}, a \textit{Structural Biologist}, and an \textit{Experimental Expert}. These agents operate within a unified reasoning context to audit the candidate mutations selected by the Rank-Stage.

\subsection{Joint Expert Review Panel}

The following prompt establishes the multi-agent persona and defines the rigorous "Check-and-Verify" protocol used to validate or reject candidates. The system enforces a strict "respect for the baseline" policy, requiring strong multi-dimensional evidence to overturn the ensemble's statistical selection.

\begin{promptbox}{System Prompt: Expert Review Panel}
Updated upon paper acceptance.
\end{promptbox}

\paragraph{Parameter Explanations:}
\begin{itemize}
    \item \placeholder{top\_k}: The target number of mutations to select (typically 30 for wet-lab validation).
    \item \texttt{Model\_Ranks\_Detail}: A dynamically generated string showing the rank of a specific mutant across all expert models (e.g., "GEMME\#3; ProSST\#12").
    \item \texttt{pLDDT}: The local confidence score from AlphaFold2, used to weight the reliability of structure-based vs. sequence-based predictions.
    \item \texttt{Position\_Alternatives}: A list of other high-scoring mutations at the same residue position, allowing the agent to perform local optimization.
\end{itemize}

\section{Agent and Wet-lab Detail for \texttt{Cas12i3}}\label{app:sec:agent_cas12i3}

\subsection{Function of \texttt{Cas12i3}}
The discovery of CRISPR-Cas systems has revolutionized the field of life sciences, offering programmable tools for precise genome editing \cite{doudna2014casreview}. In these systems, a Cas nuclease is directed by a guide RNA (crRNA) to a specific DNA sequence, where it induces a double-strand break. This cleavage triggers endogenous DNA repair pathways, such as non-homologous end joining, which can be exploited to disrupt genes or introduce precise modifications. While widely adopted effectors like SpCas9 (Type II) and AsCas12a (Type V) have enabled diverse applications, their large molecular size (approximately 1,300 amino acids) often restricts their delivery via size-constrained viral vectors, such as Adeno-Associated Viruses (AAV) \cite{jinek2012programmable,zetsche2015cpf1,wu2010effect}.

\texttt{Cas12i3} is a prominent member of the Type V-I CRISPR-Cas subfamily, characterized by its naturally compact architecture \cite{yan2019functionally}. Comprising 1,045 amino acids, \texttt{Cas12i3} is significantly smaller than both SpCas9 ($\sim$1,368 aa) and AsCas12a ($\sim$1,300 aa). This reduced size provides a critical advantage for therapeutic applications, allowing for more efficient packaging and delivery in vivo compared to larger Cas orthologs \cite{duan2024cas12i3}.

Despite its structural advantages, the wild-type (WT) \texttt{Cas12i3} protein exhibits functional limitations that constrain its broad utility. First, the baseline gene editing efficiency of WT \texttt{Cas12i3} in mammalian cells is relatively low, typically around 30$\%$, which is inferior to established editors like Cas9 \cite{wang2025engineered}. Second, strictly natural Cas nucleases often require specific Protospacer Adjacent Motifs (PAMs), such as the T-rich motifs (e.g., TTN) preferred by \texttt{Cas12i3}, which limits the accessible genomic target space \cite{lv2024genome}. Recent studies have demonstrated that rational protein engineering can overcome these bottlenecks. For instance, specific mutations have been shown to significantly boost the catalytic activity of \texttt{Cas12i3} variants (e.g., Cas-SF01) and expand their targeting range to recognize non-canonical PAMs, all while maintaining high genome-wide specificity \cite{duan2024cas12i3,chen2026structure}. Therefore, developing computational models to predict potent mutations for \texttt{Cas12i3} is of high value for creating compact, high-efficiency, and broad-spectrum genome editing tools.

\subsection{Agent Selection Detail}

We present the detailed reasoning traces and decision logs from the \model~agent panel during the specific case study of \texttt{Cas12i3}. Table~\ref{app:tab:cas12i3_dataset_input} summarizes the input metadata for this target.

\begin{table}[htbp]
\centering
\caption{Details of the selected molecule, source organism, and selection methodology (\texttt{Cas12i3}).}
\label{app:tab:cas12i3_dataset_input}
\resizebox{0.8\textwidth}{!}{%
    \begin{tabular}{@{}lllll@{}}
    \toprule
    \textbf{Molecule Name} & \textbf{Source Organism} & \textbf{Taxon} & \textbf{Selection Assay} & \textbf{Selection Type} \\ \midrule
    Cas12i3 & Unknown (Metagenomics) & Prokaryote & Enzyme Activity & Flow Cytometry \\ \bottomrule
    \end{tabular}%
}
\end{table}

\noindent
The following logs illustrate the dynamic collaboration between agents. For \texttt{Cas12i3}, the \textit{Virtual Biologist} identified a critical data scarcity issue ("Orphan Protein"), which triggered the \textit{Computational Expert} to autonomously re-calibrate the ensemble weights, shifting focus from evolutionary models to structural models.

\begin{promptbox}{Agent 1 Log: Virtual Biologist Profiling (\texttt{Cas12i3})}
Updated upon paper acceptance.
\end{promptbox}

\noindent
Based on this biological profile, the \textit{Computational Expert} (powered by \textsc{DeepSeek-Reasoner}) formulated a bespoke weighting strategy to mitigate the lack of evolutionary data.

\begin{promptbox}{Agent 2 Log: Computational Expert Strategy (\texttt{Cas12i3})}
Updated upon paper acceptance.
\end{promptbox}

\noindent
Subsequently, the \textit{Senior Expert Review Panel} conducted a biophysical audit on the top candidates. For \texttt{Cas12i3}, the panel enforced a strict position-diversity constraint (maximum 2 mutations per residue) to prevent clustering, leading to 6 strategic replacements. The detailed audit trace is shown below.

\begin{promptbox}{Agent 3 Log: Expert Review Panel Audit (\texttt{Cas12i3})}
Updated upon paper acceptance.
\end{promptbox}

\subsection{Wet-lab Protocol}

We employed the EGxxFP reporter system to quantify \texttt{Cas12i3} nuclease activity, which serves as a sensitive proxy for double-strand break (DSB) induction~\cite{duan2024cas12i3,chen2026structure}. In this system, the target DNA sequence recognized by the crRNA is embedded between two truncated EGFP fragments that share overlapping homologous repeats. Upon successful cleavage by an active \texttt{Cas12i3} variant, the functional EGFP gene is reconstituted through single-strand annealing (SSA) repair mechanisms, thereby restoring cellular green fluorescence. This design enables direct fluorescence-based readout of nuclease activity at the single-cell level.

The top $N=30$ candidate mutations identified by \model~were introduced into a human codon-optimized \texttt{Cas12i3} backbone via site-directed mutagenesis. Each variant was cloned into an all-in-one expression plasmid that encodes both the \texttt{Cas12i3} protein and its corresponding crRNA under appropriate promoters for mammalian expression.

HEK293T cells were maintained in DMEM supplemented with 10\% FBS and 1\% penicillin-streptomycin at 37°C with 5\% CO$_2$. For the activity assay, cells were seeded in 12-well plates and transfected at approximately 80\% confluency. Each well received two plasmid components: (1) the all-in-one plasmid encoding the \texttt{Cas12i3} variant and crRNA, and (2) the target-specific EGxxFP reporter plasmid which constitutively expresses an mCherry marker. Transfection was performed using Lipofectamine 2000.

Forty-eight hours post-transfection, cells were harvested by trypsinization and resuspended in PBS for flow cytometry analysis. For each sample, at least 10,000 events were recorded. Cells were gated for single cells based on forward and side scatter profiles. The cleavage activity for each \texttt{Cas12i3} variant was quantified as the ratio of EGFP-positive cells to mCherry-positive cells ($N_{\text{GFP}+}/N_{\text{mCherry}+}$). This normalization accounts for variations in transfection efficiency across samples. To facilitate direct comparison across independent experimental batches, all activity scores were further normalized relative to the Wild-Type (WT) \texttt{Cas12i3} baseline, which was included in every experiment as an internal standard.

\section{Additional Experiments}\label{app:sec:additional_exp}

\begin{table}[htbp]
\centering
\caption{Performance comparison (Spearman correlation) across varying MSA depths and Taxon categories. \textbf{Seq.}: Sequence; \textbf{Struct.}: Structure; \textbf{LLM}: Large Language Model inference.}
\label{app:tab:add_proteingym}
\resizebox{\textwidth}{!}{%
\begin{tabular}{@{}lccccccccccc@{}}
    \toprule
    \multirow{2}{*}{\textbf{Model}} & \multicolumn{4}{c}{\textbf{Inputs}} & \multicolumn{3}{c}{\textbf{Spearman by MSA Depth}} & \multicolumn{4}{c}{\textbf{Spearman by Taxon}} \\ \cmidrule(lr){2-5} \cmidrule(lr){6-8} \cmidrule(l){9-12}
     & \textbf{Seq.} & \textbf{Struct.} & \textbf{MSA} & \textbf{LLM} & \textbf{Low} & \textbf{Medium} & \textbf{High} & \textbf{Human} & \textbf{Eukaryote} & \textbf{Prokaryote} & \textbf{Virus} \\ \midrule
    \textsc{ProGen3-3B} & \y & \n & \n & \n & 0.325 & 0.407 & 0.453 & 0.410 & 0.433 & 0.392 & 0.414 \\
    \textsc{ESM-C} & \y & \n & \n & \n & 0.337 & 0.399 & 0.520 & 0.468 & 0.481 & 0.441 & 0.242 \\
    \textsc{ESM2-650M} & \y & \n & \n & \n & 0.336 & 0.423 & 0.485 & 0.442 & 0.477 & 0.458 & 0.294 \\ 
    \midrule
    \textsc{ProtSSN-Ensemble} & \y & \y & \n & \n & 0.409 & 0.454 & 0.524 & 0.470 & 0.528 & 0.492 & 0.370 \\
    \textsc{SaProt-AF650M} & \y & \y & \n & \n & 0.394 & 0.446 & 0.546 & 0.478 & 0.529 & 0.514 & 0.320 \\
    \textsc{ProSST-2048} & \y & \y & \n & \n & 0.465 & 0.507 & 0.580 & 0.516 & 0.573 & 0.549 & 0.454 \\ 
    \midrule
    \textsc{MSA-Transformer} & \y & \n & \y & \n & 0.375 & 0.456 & 0.479 & 0.439 & 0.516 & 0.446 & 0.421 \\
    \textsc{GEMME} & \y & \n & \y & \n & 0.446 & 0.474 & 0.493 & 0.469 & 0.516 & 0.467 & 0.472 \\
    \textsc{TranceptEVE-L} & \y & \n & \y & \n & 0.436 & 0.472 & 0.490 & 0.473 & 0.513 & 0.455 & 0.461 \\ 
    \midrule
    \textsc{S3F-MSA} & \y & \y & \y & \n & 0.469 & 0.509 & 0.547 & 0.502 & 0.558 & 0.521 & 0.502 \\
    \textsc{AIDO-Protein-RAG}* & \y & \y & \y & \n & 0.498 & 0.534 & 0.585 & 0.531 & 0.587 & 0.558 & 0.522 \\
    \textsc{VenusREM}* & \y & \y & \y & \n & 0.495 & 0.524 & 0.577 & 0.529 & 0.582 & 0.549 & 0.492 \\ 
    \midrule
    \textsc{Deepseek-Reasoner} & \y & \n & \n & \y & 0.179 & 0.166 & 0.164 & 0.164 & 0.190 & 0.165 & 0.160 \\ 
    \midrule
    \model-Ensemble & \y & \y & \y & \n & 0.502 & 0.547 & \textbf{0.606} & 0.550 & 0.604 & 0.576 & 0.513 \\
    \model-Rank-Stage & \y & \y & \y & \y & \textbf{0.519} & \textbf{0.562} & 0.602 & \textbf{0.556} & \textbf{0.605} & \textbf{0.583} & \textbf{0.537} \\ \bottomrule
\end{tabular}%
}
\end{table}

\begin{table}[htbp]
\centering
\caption{Rank-Stage performance comparison (Spearman correlation) of different LLM backbones across Taxon and Functional categories.}
\label{app:tab:llm_backbone_comparison}
\resizebox{\textwidth}{!}{%
    \begin{tabular}{@{}lcccccccccc@{}}
    \toprule
    \multirow{2}{*}{\textbf{LLM}} & \multirow{2}{*}{\textbf{Avg. Spearman}} & \multicolumn{4}{c}{\textbf{Spearman by Taxon}} & \multicolumn{5}{c}{\textbf{Spearman by Function}} \\ 
    \cmidrule(lr){3-6} \cmidrule(l){7-11}
     &  & \textbf{Human} & \textbf{Eukaryote} & \textbf{Prokaryote} & \textbf{Virus} & \textbf{Activity} & \textbf{Binding} & \textbf{Expression} & \textbf{Org. Fitness} & \textbf{Stability} \\ \midrule
    Ensemble (w/o LLM) & 0.542 & 0.550 & 0.604 & 0.576 & 0.513 & 0.533 & 0.478 & 0.485 & 0.556 & 0.661 \\
    \textsc{Qwen3-8B} & 0.543 & 0.551 & 0.603 & 0.574 & 0.525 & 0.532 & 0.473 & 0.492 & 0.554 & \textbf{0.663} \\
    \textsc{GLM-4.5-Flash} & 0.548 & 0.555 & \textbf{0.607} & 0.577 & \textbf{0.550} & \textbf{0.540} & 0.477 & 0.509 & 0.556 & 0.660 \\
    \textsc{GPT-5-mini} & 0.548 & \textbf{0.553} & 0.602 & 0.577 & 0.554 & \textbf{0.541} & 0.479 & \textbf{0.511} & \textbf{0.558} & 0.651 \\
    \textsc{Deepseek-Chat} & 0.549 & \textbf{0.556} & 0.606 & 0.578 & 0.546 & 0.540 & 0.480 & 0.509 & \textbf{0.558} & 0.658 \\
    \textsc{Deepseek-Reasoner} & \textbf{0.551} & \textbf{0.556} & 0.605 & \textbf{0.583} & 0.537 & 0.539 & \textbf{0.497} & 0.505 & \textbf{0.558} & 0.658 \\
     \bottomrule
    \end{tabular}%
}
\end{table}

\begin{table}[htbp]
\centering
\caption{Rank-Stage stability analysis: Number of error files and Average Spearman correlation across 3 independent runs.}
\label{app:tab:llm_multi_run}
\resizebox{0.5\textwidth}{!}{%
    \begin{tabular}{@{}lccc@{}}
    \toprule
    \textbf{LLM} & \textbf{Run} & \textbf{\# Error File} & \textbf{Avg. Spearman} \\ \midrule
    \multirow{3}{*}{\textsc{Deepseek-Reasoner}} & 1 & 2 & 0.551 \\
     & 2 & 1 & 0.550 \\
     & 3 & 1 & 0.551 \\ \midrule
    \multirow{3}{*}{\textsc{GPT-5-mini}} & 1 & 7 & 0.548 \\
     & 2 & 4 & 0.547 \\
     & 3 & 8 & 0.549 \\ \bottomrule
    \end{tabular}%
}
\end{table}

\begin{table}[htbp]
\centering
\caption{Performance comparison (\%) on different LLMs under different stages and selection budgets ($N$).
The format is reported as $Mean_{(Std)}$. \textbf{Rank-Stage}: Top ${N}$ mutants of LLM-Ensembled scores depending on specific PLMs' scores; \textbf{Reason-Stage}: LLM-selected mutants according to multiple score-sets given from PLM, \model-Ensemble and \model-Rank-Stage.}
\label{app:tab:rar_budget_full}
\resizebox{\textwidth}{!}{%
\begin{tabular}{@{}c l l ccccccc@{}}
    \toprule
    \textbf{Budget} & \textbf{LLM} & \textbf{Stage} &
    \textbf{Max} & \textbf{Top 1\%} & \textbf{Top 5\%} & \textbf{Top 10\%} &
    \textbf{Hit 30} & \textbf{Hit 10} & \textbf{Hit 5} \\
    \midrule
    
    \multirow{11}{*}{40}
    & Ensemble & -- &
    65.78 & 1.85 & 11.77 & 23.95 &
    3.76 & 1.94 & 1.29 \\
    \cmidrule(l){2-10}
    
    & \multirow{2}{*}{\textsc{GPT-5-mini}}
    & Rank-Stage &
    66.79$_{(0.53)}$ & 1.93$_{(0.22)}$ & 12.28$_{(0.45)}$ & 22.72$_{(0.53)}$ &
    4.41$_{(0.29)}$ & 3.55$_{(0.32)}$ & 4.52$_{(0.65)}$ \\
    & & Reason-Stage &
    67.84$_{(0.64)}$ & 2.18$_{(0.16)}$ & 12.15$_{(0.26)}$ & 22.58$_{(0.29)}$ &
    4.52$_{(0.29)}$ & 4.08$_{(0.18)}$ & 4.09$_{(0.38)}$ \\
    \cmidrule(l){2-10}
    
    & \multirow{2}{*}{\textsc{GLM-4.5-Flash}}
    & Rank-Stage &
    66.14$_{(1.24)}$ & 1.91$_{(0.26)}$ & 11.77$_{(0.61)}$ & 21.78$_{(0.69)}$ &
    4.57$_{(0.34)}$ & 3.53$_{(0.23)}$ & 4.11$_{(0.71)}$ \\
    & & Reason-Stage &
    66.61$_{(0.02)}$ & 1.87$_{(0.04)}$ & 11.21$_{(0.07)}$ & 21.11$_{(0.04)}$ &
    4.57$_{(0.11)}$ & 2.69$_{(0.43)}$ & 4.52$_{(0.00)}$ \\
    \cmidrule(l){2-10}
    
    & \multirow{2}{*}{\textsc{Qwen3-8B}}
    & Rank-Stage &
    65.71$_{(0.91)}$ & 1.69$_{(0.23)}$ & 11.41$_{(0.81)}$ & 21.79$_{(1.46)}$ &
    4.08$_{(0.34)}$ & 2.40$_{(0.33)}$ & 2.15$_{(0.29)}$ \\
    & & Reason-Stage &
    67.24$_{(1.82)}$ & 2.02$_{(0.41)}$ & 11.74$_{(0.66)}$ & 22.18$_{(0.76)}$ &
    4.17$_{(0.45)}$ & 2.90$_{(0.87)}$ & 3.23$_{(1.29)}$ \\
    \cmidrule(l){2-10}
    
    & \multirow{2}{*}{\textsc{Deepseek-Chat}}
    & Rank-Stage &
    67.30$_{(0.12)}$ & 2.07$_{(0.09)}$ & 12.36$_{(0.60)}$ & 23.20$_{(0.33)}$ &
    4.48$_{(0.06)}$ & 3.76$_{(0.18)}$ & 4.73$_{(0.37)}$ \\
    & & Reason-Stage &
    67.67$_{(0.40)}$ & 2.23$_{(0.12)}$ & 12.63$_{(0.30)}$ & 23.76$_{(0.38)}$ &
    4.80$_{(0.27)}$ & 3.98$_{(0.37)}$ & 4.73$_{(0.37)}$ \\
    \cmidrule(l){2-10}
    
    & \multirow{2}{*}{\textsc{Deepseek-Reasoner}}
    & Rank-Stage &
    68.42$_{(1.17)}$ & 2.15$_{(0.05)}$ & 12.66$_{(0.32)}$ & 23.55$_{(0.37)}$ &
    4.84$_{(0.18)}$ & 4.41$_{(0.38)}$ & 5.38$_{(0.38)}$ \\
    & & Reason-Stage &
    69.50$_{(2.17)}$ & 2.31$_{(0.24)}$ & 12.10$_{(0.49)}$ & 23.15$_{(0.17)}$ &
    5.05$_{(0.49)}$ & 4.84$_{(0.65)}$ & 4.73$_{(0.99)}$ \\
    
    \midrule
    
    \multirow{11}{*}{30}
    & Ensemble & -- &
    62.22 & 1.51 & 11.72 & 22.80 &
    3.01 & 1.61 & 1.29 \\
    \cmidrule(l){2-10}
    
    & \multirow{2}{*}{\textsc{GPT-5-mini}}
    & Rank-Stage &
    65.89$_{(0.16)}$ & 2.22$_{(0.06)}$ & 12.65$_{(0.17)}$ & 23.05$_{(0.34)}$ &
    3.76$_{(0.00)}$ & 3.12$_{(0.19)}$ & 3.66$_{(0.37)}$ \\
    & & Reason-Stage &
    65.96$_{(0.22)}$ & 2.33$_{(0.22)}$ & 13.08$_{(0.13)}$ & 23.19$_{(0.06)}$ &
    3.91$_{(0.13)}$ & 3.34$_{(0.18)}$ & 3.87$_{(0.00)}$ \\
    \cmidrule(l){2-10}
    
    & \multirow{2}{*}{\textsc{GLM-4.5-Flash}}
    & Rank-Stage &
    64.17$_{(1.23)}$ & 2.00$_{(0.11)}$ & 12.12$_{(0.90)}$ & 22.10$_{(1.00)}$ &
    3.75$_{(0.12)}$ & 2.79$_{(0.31)}$ & 3.45$_{(0.30)}$ \\
    & & Reason-Stage &
    65.23$_{(0.68)}$ & 1.86$_{(0.11)}$ & 11.05$_{(0.26)}$ & 20.75$_{(0.09)}$ &
    3.69$_{(0.05)}$ & 2.74$_{(0.18)}$ & 3.55$_{(0.37)}$ \\
    \cmidrule(l){2-10}
    
    & \multirow{2}{*}{\textsc{Qwen3-8B}}
    & Rank-Stage &
    64.53$_{(1.39)}$ & 1.88$_{(0.33)}$ & 12.00$_{(0.74)}$ & 22.00$_{(1.02)}$ &
    3.31$_{(0.22)}$ & 2.15$_{(0.58)}$ & 1.99$_{(0.53)}$ \\
    & & Reason-Stage &
    66.58$_{(0.43)}$ & 2.34$_{(0.22)}$ & 12.74$_{(0.19)}$ & 22.77$_{(0.28)}$ &
    3.74$_{(0.26)}$ & 2.66$_{(0.16)}$ & 2.91$_{(0.38)}$ \\
    \cmidrule(l){2-10}
    
    & \multirow{2}{*}{\textsc{Deepseek-Chat}}
    & Rank-Stage &
    65.87$_{(0.30)}$ & 2.33$_{(0.06)}$ & 13.08$_{(0.22)}$ & 23.25$_{(0.07)}$ &
    3.76$_{(0.11)}$ & 3.12$_{(0.19)}$ & 3.66$_{(0.37)}$ \\
    & & Reason-Stage &
    65.92$_{(0.17)}$ & 2.40$_{(0.12)}$ & 12.98$_{(0.51)}$ & 23.44$_{(0.75)}$ &
    3.80$_{(0.16)}$ & 3.34$_{(0.18)}$ & 3.87$_{(0.00)}$ \\
    \cmidrule(l){2-10}
    
    & \multirow{2}{*}{\textsc{Deepseek-Reasoner}}
    & Rank-Stage &
    66.25$_{(0.40)}$ & 2.33$_{(0.22)}$ & 13.24$_{(0.36)}$ & 23.33$_{(0.56)}$ &
    4.09$_{(0.18)}$ & 3.76$_{(0.49)}$ & 4.09$_{(0.38)}$ \\
    & & Reason-Stage &
    68.42$_{(1.70)}$ & 2.40$_{(0.06)}$ & 13.19$_{(0.53)}$ & 22.65$_{(0.61)}$ &
    4.41$_{(0.11)}$ & 4.19$_{(0.33)}$ & 4.52$_{(0.65)}$ \\
    
    \midrule
    
    \multirow{11}{*}{20}
    & Ensemble & -- &
    59.75 & 2.26 & 12.58 & 25.48 &
    2.37 & 1.61 & 1.29 \\
    \cmidrule(l){2-10}
    
    & \multirow{2}{*}{\textsc{GPT-5-mini}}
    & Rank-Stage &
    63.18$_{(0.83)}$ & 2.53$_{(0.33)}$ & 12.85$_{(0.18)}$ & 22.74$_{(0.28)}$ &
    2.62$_{(0.13)}$ & 2.58$_{(0.32)}$ & 3.01$_{(0.38)}$ \\
    & & Reason-Stage &
    64.14$_{(0.04)}$ & 2.63$_{(0.18)}$ & 14.09$_{(0.38)}$ & 23.77$_{(0.67)}$ &
    2.62$_{(0.06)}$ & 2.58$_{(0.55)}$ & 3.44$_{(0.74)}$ \\
    \cmidrule(l){2-10}
    
    & \multirow{2}{*}{\textsc{GLM-4.5-Flash}}
    & Rank-Stage &
    61.05$_{(0.68)}$ & 2.13$_{(0.27)}$ & 12.42$_{(1.20)}$ & 22.10$_{(1.56)}$ &
    2.57$_{(0.26)}$ & 2.14$_{(0.24)}$ & 2.31$_{(0.43)}$ \\
    & & Reason-Stage &
    61.12$_{(0.55)}$ & 1.90$_{(0.16)}$ & 11.46$_{(0.57)}$ & 20.24$_{(0.60)}$ &
    2.34$_{(0.24)}$ & 2.10$_{(0.32)}$ & 2.10$_{(0.32)}$ \\
    \cmidrule(l){2-10}
    
    & \multirow{2}{*}{\textsc{Qwen3-8B}}
    & Rank-Stage &
    62.47$_{(1.66)}$ & 2.11$_{(0.43)}$ & 13.00$_{(1.20)}$ & 23.54$_{(1.02)}$ &
    2.54$_{(0.13)}$ & 1.65$_{(0.37)}$ & 1.31$_{(0.51)}$ \\
    & & Reason-Stage &
    64.04$_{(0.09)}$ & 2.42$_{(0.00)}$ & 14.30$_{(0.25)}$ & 24.52$_{(0.17)}$ &
    2.55$_{(0.16)}$ & 1.94$_{(0.00)}$ & 1.94$_{(0.00)}$ \\
    \cmidrule(l){2-10}
    
    & \multirow{2}{*}{\textsc{Deepseek-Chat}}
    & Rank-Stage &
    63.56$_{(0.98)}$ & 2.58$_{(0.16)}$ & 13.33$_{(0.81)}$ & 23.44$_{(0.73)}$ &
    2.72$_{(0.27)}$ & 2.69$_{(0.18)}$ & 3.01$_{(0.38)}$ \\
    & & Reason-Stage &
    63.76$_{(1.08)}$ & 2.58$_{(0.28)}$ & 13.34$_{(0.09)}$ & 24.03$_{(0.32)}$ &
    2.44$_{(0.16)}$ & 2.58$_{(0.65)}$ & 3.01$_{(0.74)}$ \\
    \cmidrule(l){2-10}
    
    & \multirow{2}{*}{\textsc{Deepseek-Reasoner}}
    & Rank-Stage &
    64.19$_{(0.87)}$ & 2.47$_{(0.24)}$ & 13.87$_{(0.58)}$ & 23.71$_{(0.42)}$ &
    2.58$_{(0.38)}$ & 2.80$_{(0.49)}$ & 3.12$_{(0.19)}$ \\
    & & Reason-Stage &
    64.30$_{(0.58)}$ & 2.58$_{(0.13)}$ & 14.03$_{(0.74)}$ & 24.19$_{(0.73)}$ &
    2.72$_{(0.16)}$ & 2.50$_{(0.16)}$ & 3.39$_{(0.62)}$ \\
    
    \midrule
    
    \multirow{11}{*}{10}
    & Ensemble & -- &
    57.12 & 1.94 & 12.58 & 24.19 &
    1.29 & 0.97 & 0.65 \\
    \cmidrule(l){2-10}
    
    & \multirow{2}{*}{\textsc{GPT-5-mini}}
    & Rank-Stage &
    58.20$_{(0.72)}$ & 2.47$_{(0.18)}$ & 13.76$_{(1.22)}$ & 24.09$_{(0.81)}$ &
    1.58$_{(0.12)}$ & 1.29$_{(0.00)}$ & 1.29$_{(0.00)}$ \\
    & & Reason-Stage &
    59.50$_{(0.36)}$ & 3.12$_{(0.19)}$ & 15.38$_{(0.81)}$ & 24.28$_{(1.30)}$ &
    1.68$_{(0.06)}$ & 1.94$_{(0.00)}$ & 2.15$_{(0.74)}$ \\
    \cmidrule(l){2-10}
    
    & \multirow{2}{*}{\textsc{GLM-4.5-Flash}}
    & Rank-Stage &
    57.62$_{(0.72)}$ & 3.12$_{(0.27)}$ & 14.49$_{(1.47)}$ & 24.67$_{(2.45)}$ &
    1.67$_{(0.11)}$ & 1.56$_{(0.19)}$ & 1.81$_{(0.35)}$ \\
    & & Reason-Stage &
    58.99$_{(0.57)}$ & 3.47$_{(0.16)}$ & 12.99$_{(1.16)}$ & 21.70$_{(1.27)}$ &
    1.64$_{(0.22)}$ & 2.18$_{(0.16)}$ & 2.42$_{(0.32)}$ \\
    \cmidrule(l){2-10}
    
    & \multirow{2}{*}{\textsc{Qwen3-8B}}
    & Rank-Stage &
    58.38$_{(1.24)}$ & 2.56$_{(0.54)}$ & 15.41$_{(2.32)}$ & 25.58$_{(2.40)}$ &
    1.40$_{(0.27)}$ & 1.16$_{(0.33)}$ & 1.00$_{(0.34)}$ \\
    & & Reason-Stage &
    59.51$_{(0.50)}$ & 3.15$_{(0.31)}$ & 16.94$_{(0.85)}$ & 27.90$_{(0.56)}$ &
    1.59$_{(0.05)}$ & 1.45$_{(0.18)}$ & 1.61$_{(0.65)}$ \\
    \cmidrule(l){2-10}
    
    & \multirow{2}{*}{\textsc{Deepseek-Chat}}
    & Rank-Stage &
    58.71$_{(0.24)}$ & 2.37$_{(0.18)}$ & 12.80$_{(0.38)}$ & 24.09$_{(0.38)}$ &
    1.51$_{(0.00)}$ & 1.08$_{(0.18)}$ & 0.86$_{(0.37)}$ \\
    & & Reason-Stage &
    60.24$_{(0.68)}$ & 3.23$_{(0.00)}$ & 14.62$_{(0.67)}$ & 24.95$_{(0.37)}$ &
    1.72$_{(0.00)}$ & 1.83$_{(0.18)}$ & 2.37$_{(0.37)}$ \\
    \cmidrule(l){2-10}
    
    & \multirow{2}{*}{\textsc{Deepseek-Reasoner}}
    & Rank-Stage &
    58.14$_{(0.37)}$ & 2.75$_{(0.16)}$ & 14.95$_{(0.37)}$ & 26.13$_{(0.00)}$ &
    1.69$_{(0.16)}$ & 1.40$_{(0.18)}$ & 1.29$_{(0.00)}$ \\
    & & Reason-Stage &
    58.77$_{(1.04)}$ & 3.12$_{(0.19)}$ & 14.62$_{(0.38)}$ & 24.95$_{(0.99)}$ &
    1.61$_{(0.11)}$ & 1.94$_{(0.00)}$ & 2.58$_{(0.00)}$ \\
    
    \bottomrule
\end{tabular}}
\end{table}

\begin{figure*}[t]
  \vskip 0.2in
  \begin{center}
    \centerline{\includegraphics[width=0.6\textwidth]{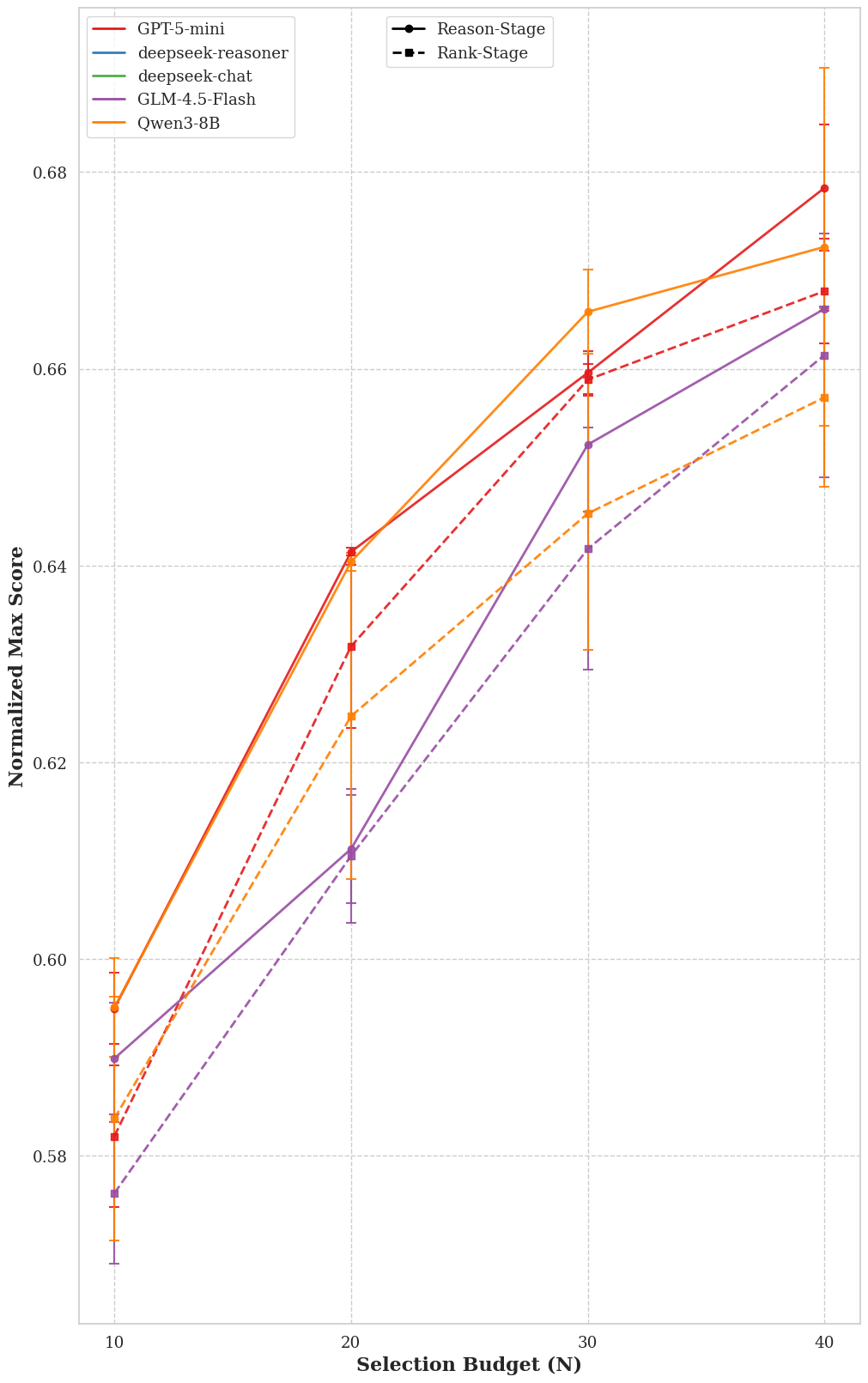}}
    \caption{
      Exploration frontier of Normalized Max Score vs. Selection Budget ($N$). Solid lines represent Reason-Stage, and dashed lines represent Rank-Stage. Error bars denote standard deviation. Stronger models (e.g., \textsc{Deepseek-Reasoner}) demonstrate higher exploration efficiency.
    }
    \label{app:fig:max_score}
  \end{center}
\end{figure*}

\begin{table}[htbp]
\centering
\caption{Performance comparison (\%) of mutation selection solely on LLMs under different budgets ($N$) and varying zero-shot settings. The format is reported as $Mean_{(Std)}$. \textbf{Seq.}: Amino acid sequence; \textbf{Str.}: Protein structure; \textbf{Ann}: Protein annotation information. The inputs are directly added to the context of LLMs.}
\label{app:tab:llm_zero_shot_full}
\resizebox{\textwidth}{!}{%
\begin{tabular}{@{}c ll ccccccc@{}}
    \toprule
    \textbf{Budget} & \textbf{LLM} & \textbf{Inputs} & \textbf{Max} & \textbf{Top 1\%} & \textbf{Top 5\%} & \textbf{Top 10\%} & \textbf{Hit 5} & \textbf{Hit 10} & \textbf{Hit 30} \\ \midrule
    
    \multirow{7}{*}{40} 
     & Random & - & $\res{61.09}{1.78}$ & $\res{0.78}{0.25}$ & $\res{4.97}{0.79}$ & $\res{9.97}{0.81}$ & $\res{1.55}{0.39}$ & $\res{1.44}{0.84}$ & $\res{1.93}{0.17}$ \\ 
     
     \cmidrule(l){2-10}
     & \multirow{3}{*}{\textsc{GPT-5-mini}} & Seq. & $\res{59.15}{0.63}$ & $\res{0.86}{0.30}$ & $\res{5.61}{0.63}$ & $\res{12.07}{0.80}$ & $\res{0.67}{0.00}$ & $\res{1.11}{0.38}$ & $\res{1.48}{0.07}$ \\
     & & Seq.+Ann. & $\res{61.02}{1.10}$ & $\res{1.33}{0.30}$ & $\res{7.42}{0.14}$ & $\res{14.19}{0.27}$ & $\res{1.78}{1.02}$ & $\res{1.44}{0.51}$ & $\res{2.07}{0.46}$ \\
     & & Seq.+Ann.+Str. & $\res{61.50}{0.42}$ & $\res{1.19}{0.13}$ & $\res{7.78}{0.84}$ & $\res{15.58}{0.88}$ & $\res{1.55}{0.39}$ & $\res{1.22}{0.39}$ & $\res{2.04}{0.25}$ \\ 
     
     \cmidrule(l){2-10}
     & \multirow{3}{*}{\textsc{DeepSeek-Reasoner}} & Seq. & $\res{59.05}{0.58}$ & $\res{0.69}{0.10}$ & $\res{5.80}{0.72}$ & $\res{12.83}{1.04}$ & $\res{1.11}{0.38}$ & $\res{1.22}{0.69}$ & $\res{1.44}{0.22}$ \\
     & & Seq.+Ann. & $\res{60.79}{0.83}$ & $\res{1.06}{0.20}$ & $\res{6.20}{0.67}$ & $\res{12.83}{0.51}$ & $\res{2.22}{0.39}$ & $\res{1.67}{0.34}$ & $\res{1.85}{0.17}$ \\
     & & Seq.+Ann.+Str. & $\res{60.59}{1.00}$ & $\res{0.94}{0.27}$ & $\res{7.09}{0.88}$ & $\res{15.59}{0.66}$ & $\res{1.78}{0.77}$ & $\res{1.78}{1.07}$ & $\res{2.07}{0.46}$ \\ 
     \midrule
    
    \multirow{7}{*}{30} 
     & Random & - & $\res{58.41}{3.40}$ & $\res{0.85}{0.28}$ & $\res{4.41}{0.67}$ & $\res{9.08}{0.45}$ & $\res{1.33}{0.67}$ & $\res{1.11}{0.38}$ & $\res{0.96}{0.13}$ \\ 
     \cmidrule(l){2-10}
     & \multirow{3}{*}{\textsc{GPT-5-mini}} & Seq. & $\res{58.17}{1.90}$ & $\res{0.85}{0.28}$ & $\res{6.37}{0.67}$ & $\res{13.37}{1.23}$ & $\res{0.67}{0.67}$ & $\res{0.78}{0.51}$ & $\res{1.07}{0.17}$ \\
     & & Seq.+Ann. & $\res{57.72}{0.90}$ & $\res{0.71}{0.06}$ & $\res{6.74}{0.28}$ & $\res{14.11}{0.22}$ & $\res{0.67}{0.67}$ & $\res{0.67}{0.67}$ & $\res{1.33}{0.23}$ \\
     & & Seq.+Ann.+Str. & $\res{59.00}{1.23}$ & $\res{1.15}{0.45}$ & $\res{7.37}{0.74}$ & $\res{14.48}{0.45}$ & $\res{0.45}{0.39}$ & $\res{0.89}{0.51}$ & $\res{1.52}{0.34}$ \\ 
     \cmidrule(l){2-10}
     & \multirow{3}{*}{\textsc{DeepSeek-Reasoner}} & Seq. & $\res{55.52}{2.87}$ & $\res{0.70}{0.23}$ & $\res{5.78}{0.62}$ & $\res{11.70}{0.50}$ & $\res{0.67}{0.67}$ & $\res{0.89}{0.38}$ & $\res{1.15}{0.23}$ \\
     & & Seq.+Ann. & $\res{60.12}{1.77}$ & $\res{1.15}{0.55}$ & $\res{7.11}{1.26}$ & $\res{13.97}{2.45}$ & $\res{1.55}{0.39}$ & $\res{1.22}{0.51}$ & $\res{1.67}{0.20}$ \\
     & & Seq.+Ann.+Str. & $\res{57.61}{1.55}$ & $\res{0.78}{0.51}$ & $\res{6.92}{0.57}$ & $\res{15.18}{0.28}$ & $\res{0.89}{0.77}$ & $\res{0.56}{0.20}$ & $\res{1.37}{0.13}$ \\ 
     \midrule
    
    \multirow{7}{*}{20} 
     & Random & - & $\res{54.11}{2.03}$ & $\res{0.78}{0.48}$ & $\res{4.83}{0.50}$ & $\res{8.83}{0.44}$ & $\res{0.45}{0.39}$ & $\res{0.78}{0.51}$ & $\res{1.04}{0.06}$ \\ 
     \cmidrule(l){2-10}
     & \multirow{3}{*}{\textsc{GPT-5-mini}} & Seq. & $\res{53.45}{1.82}$ & $\res{0.67}{0.34}$ & $\res{6.33}{0.44}$ & $\res{12.11}{0.82}$ & $\res{0.00}{0.00}$ & $\res{0.22}{0.39}$ & $\res{0.74}{0.13}$ \\
     & & Seq.+Ann. & $\res{55.68}{0.59}$ & $\res{1.06}{0.51}$ & $\res{7.72}{0.25}$ & $\res{15.44}{0.79}$ & $\res{1.11}{0.77}$ & $\res{0.89}{0.38}$ & $\res{1.33}{0.19}$ \\
     & & Seq.+Ann.+Str. & $\res{55.65}{1.27}$ & $\res{0.89}{0.38}$ & $\res{6.89}{1.09}$ & $\res{14.94}{0.19}$ & $\res{0.00}{0.00}$ & $\res{0.33}{0.00}$ & $\res{0.93}{0.17}$ \\ 
     \cmidrule(l){2-10}
     & \multirow{3}{*}{\textsc{DeepSeek-Reasoner}} & Seq. & $\res{52.49}{2.59}$ & $\res{0.39}{0.19}$ & $\res{5.22}{0.58}$ & $\res{10.84}{1.09}$ & $\res{0.22}{0.39}$ & $\res{0.22}{0.19}$ & $\res{0.74}{0.23}$ \\
     & & Seq.+Ann. & $\res{55.66}{1.33}$ & $\res{1.22}{0.35}$ & $\res{6.94}{0.42}$ & $\res{14.72}{0.54}$ & $\res{1.11}{0.38}$ & $\res{1.34}{0.58}$ & $\res{1.15}{0.13}$ \\
     & & Seq.+Ann.+Str. & $\res{56.00}{0.70}$ & $\res{1.06}{0.34}$ & $\res{7.56}{1.69}$ & $\res{16.28}{2.00}$ & $\res{0.22}{0.39}$ & $\res{0.45}{0.39}$ & $\res{1.15}{0.28}$ \\ 
     \midrule
    
    \multirow{7}{*}{10} 
     & Random & - & $\res{49.25}{3.48}$ & $\res{1.22}{0.51}$ & $\res{5.00}{1.00}$ & $\res{11.00}{2.18}$ & $\res{0.22}{0.39}$ & $\res{0.56}{0.51}$ & $\res{0.29}{0.17}$ \\ 
     \cmidrule(l){2-10}
     & \multirow{3}{*}{\textsc{GPT-5-mini}} & Seq. & $\res{48.71}{0.79}$ & $\res{0.33}{0.58}$ & $\res{5.67}{0.88}$ & $\res{13.11}{0.38}$ & $\res{0.00}{0.00}$ & $\res{0.00}{0.00}$ & $\res{0.41}{0.17}$ \\
     & & Seq.+Ann. & $\res{48.80}{1.60}$ & $\res{0.89}{0.96}$ & $\res{6.67}{0.88}$ & $\res{15.78}{1.26}$ & $\res{0.33}{0.34}$ & $\res{0.33}{0.34}$ & $\res{0.63}{0.13}$ \\
     & & Seq.+Ann.+Str. & $\res{51.25}{1.94}$ & $\res{1.44}{0.38}$ & $\res{8.11}{0.50}$ & $\res{15.87}{0.54}$ & $\res{0.22}{0.39}$ & $\res{0.22}{0.19}$ & $\res{0.44}{0.12}$ \\ 
     \cmidrule(l){2-10}
     & \multirow{3}{*}{\textsc{DeepSeek-Reasoner}} & Seq. & $\res{49.53}{2.87}$ & $\res{0.55}{0.39}$ & $\res{7.22}{0.51}$ & $\res{14.55}{2.41}$ & $\res{0.22}{0.39}$ & $\res{0.22}{0.19}$ & $\res{0.48}{0.17}$ \\
     & & Seq.+Ann. & $\res{48.62}{3.67}$ & $\res{1.11}{1.02}$ & $\res{7.78}{0.38}$ & $\res{14.00}{1.45}$ & $\res{0.89}{1.02}$ & $\res{0.67}{0.88}$ & $\res{0.78}{0.22}$ \\
     & & Seq.+Ann.+Str. & $\res{52.52}{2.37}$ & $\res{1.11}{0.38}$ & $\res{8.67}{1.76}$ & $\res{16.89}{2.21}$ & $\res{0.22}{0.39}$ & $\res{0.33}{0.34}$ & $\res{0.82}{0.13}$ \\ 
     \bottomrule
\end{tabular}%
}
\end{table}

\begin{table}[t]
\centering
\caption{Spearman correlation of different LLMs under varying Few-shot settings and Mutant Numbers.}
\label{app:tab:llm_few_shot}
\resizebox{0.6\linewidth}{!}{%
\begin{tabular}{@{}clccccc@{}}
    \toprule
    \multirow{2}{*}{\textbf{Few-shot}} & \multirow{2}{*}{\textbf{LLM}} & \multicolumn{5}{c}{\textbf{Evaluate Mutant Number (Spearman)}} \\ \cmidrule(l){3-7} 
     &  & \textbf{10} & \textbf{20} & \textbf{30} & \textbf{50} & \textbf{100} \\ 
     \midrule
    \multirow{3}{*}{10} & \textsc{GPT-5-mini} & 0.483 & 0.435 & 0.455 & 0.441 & 0.390 \\
     & \textsc{DeepSeek-Reasoner} & 0.304 & 0.305 & 0.331 & 0.295 & 0.300 \\
     & \textsc{GLM-4.5-Flash} & 0.204 & 0.234 & 0.197 & 0.243 & 0.160 \\ 
     \midrule
    \multirow{3}{*}{20} & \textsc{GPT-5-mini} & 0.296 & 0.409 & 0.435 & 0.429 & 0.418 \\
     & \textsc{DeepSeek-Reasoner} & 0.291 & 0.309 & 0.323 & 0.366 & 0.352 \\
     & \textsc{GLM-4.5-Flash} & 0.296 & 0.208 & 0.319 & 0.253 & 0.206 \\ 
     \midrule
    \multirow{3}{*}{30} & \textsc{GPT-5-mini} & 0.521 & 0.493 & 0.460 & 0.452 & 0.424 \\
     & \textsc{DeepSeek-Reasoner} & 0.296 & 0.309 & 0.323 & 0.366 & 0.352 \\
     & \textsc{GLM-4.5-Flash} & 0.321 & 0.292 & 0.300 & 0.248 & 0.181 \\ 
     \bottomrule
\end{tabular}%
}
\end{table}

\end{document}